\documentclass[aps,prl,twocolumn,superscriptaddress]{revtex4-1}

\usepackage[usenames, dvipsnames]{color}
\usepackage[english]{babel}
\usepackage[T1]{fontenc}
\usepackage[latin9]{inputenc}
\usepackage{latexsym}
\usepackage{float}
\usepackage{amsmath}
\usepackage[normalem]{ulem}
\usepackage{graphicx}
\usepackage{times}   
\usepackage{esint}

\makeatletter

\@ifundefined{textcolor}{}
{%
 \definecolor{BLACK}{gray}{0}
 \definecolor{WHITE}{gray}{1}
 \definecolor{RED}{rgb}{1,0,0}
 \definecolor{GREEN}{rgb}{0,1,0}
 \definecolor{BLUE}{rgb}{0,0,1}
 \definecolor{CYAN}{cmyk}{1,0,0,0}
 \definecolor{MAGENTA}{cmyk}{0,1,0,0}
 \definecolor{YELLOW}{cmyk}{0,0,1,0}
}

\@ifundefined{date}{}{\date{}}
\AtBeginDocument{
  
}
\makeatother

\newcommand{\SAVE}[1]{}

\begin{document}
\renewcommand\abstractname{}

\title{Resonating quantum three-coloring wavefunctions for the kagome quantum antiferromagnet}
\author{Hitesh J. Changlani}
\affiliation{Department of Physics, Florida State University, Tallahassee, Florida 32306, USA}
\affiliation{Department of Physics and Astronomy, Johns Hopkins University, Baltimore, MD 21218 
and Institute for Quantum Matter, Johns Hopkins University, Baltimore, MD 21218}
\author{Sumiran Pujari}
\affiliation{Department of Physics, Indian Institute of Technology Bombay, Mumbai, MH 400076, India}
\author{Chia-Min Chung}
\affiliation{Department of Physics and Arnold Sommerfeld Center for Theoretical Physics,
Ludwig-Maximilians-Universitat Munchen, Theresienstrasse 37, 80333 Munchen, Germany}
\author{Bryan K. Clark}
\affiliation{Institute  for  Condensed  Matter  Theory  and  Department  of  Physics, University  of Illinois at Urbana-Champaign, USA}
\date{\today}

\begin{abstract}
Motivated by the recent discovery of a macroscopically degenerate exactly solvable 
point of the spin-$1/2$ $XXZ$ model for $J_z/J=-1/2$ on the kagome lattice [H.~J.~Changlani et al.~Phys. Rev. Lett 120, 117202 (2018)] -- 
a result that holds for arbitrary magnetization -- we develop an exact mapping between its exact "quantum three-coloring" wavefunctions and the characteristic 
localized and topological magnons. This map, involving "resonating two-color loops", is developed to represent exact many-body ground state wavefunctions 
for special high magnetizations. Using this map we show that these exact ground state solutions are valid for any $J_z/J \geq-1/2$. 
This demonstrates the equivalence of the ground-state wavefunction of the Ising, Heisenberg and $XY$ regimes all the way to the 
$J_z/J=-1/2$ point for these high magnetization sectors. In the hardcore bosonic language, this means that a certain class 
of exact many-body solutions, previously argued to hold for purely \emph{repulsive} interactions ($J_z \geq 0$), 
actually hold for \emph{attractive} interactions as well, 
up to a critical interaction strength. 

For the case of zero magnetization, where the ground state is not exactly known, we perform density matrix renormalization group calculations. 
Based on the calculation of the ground state energy and measurement of order parameters, we provide evidence 
for a lack of any qualitative change in the ground state on finite clusters in 
the Ising ($J_z \gg J$), Heisenberg ($J_z=J$) and $XY$ ($J_z=0$) regimes, continuing adiabatically 
to the vicinity of the macroscopically degenerate $J_z/J=-1/2$ point. 
These findings offer a framework for recent results in the literature, and also suggest that the $J_z/J=-1/2$ point 
is an unconventional quantum critical point whose vicinity may contain the key to resolving the spin-$1/2$ kagome problem. 
\end{abstract}

\maketitle

\section{Introduction}
\label{sec:intro}
Quantum frustrated magnetism presents one of the most intriguing and intricate examples of the interplay 
between spatial geometry and quantum mechanics. This results in a rich multitude of competing 
exotic phases such as valence bond solids, topological phases including several spin liquids, and magnetically 
ordered phases. Slight changes in the material composition or geometry can lead to a dramatic change in its
phase, making frustrated magnets ideal playgrounds to study quantum phase transitions. 
 
The building blocks of many of these systems are lattices of magnetic ions made from motifs of connected triangles. 
Prominent amongst these is the kagome lattice, 
a lattice of corner sharing triangles which has been intensely studied owing to its relevance to materials 
such as Herbertsmithite (a kagome lattice of Cu$^{2+}$ ions)~\cite{Nocera_Kagome_2005}. 
Experiments on Herbertsmithite~\cite{Helton_2007, Mendels_Bert}-- of which the idealized kagome 
Heisenberg antiferromagnet is known to be a good model~\cite{Jeschke_2013} -- find that spins do not order even 
at the lowest investigated temperatures (50 mK, a small fraction of the exchange energy of 200 K), 
tantalizingly suggesting the picture of a two-dimensional spin-liquid ground state. However, in spite of 
several theoretical efforts devoted to the idealized model, there is no universal consensus on the precise nature of 
the spin liquid ground state~\cite{ZengElser, Wen_Kagome, White_Kagome, Depenbrock_Kagome, Iqbal_Kagome, Jiang_Balents, Tay_Motrunich, 
He_Zaletel_Kagome, Normand_Xiang, Messio, Hao_Tchernyshyov} and recent work even suggests that larger lattices should stabilize an ordered state~\cite{Ralko_Mila}. 
To reconcile some of these observations, it has been suggested that the kagome Heisenberg model lies at or 
close to a critical point in the phase diagram in a suitably chosen parameter space of model Hamiltonians~\cite{Changlani_PRL, Tao_Li_critical1}. 

Previous work (by two of us, HJC and BKC in collaboration with others) 
contributed to the understanding of the kagome phase diagram through the discovery of 
an extensively \textit{quantum} degenerate exactly solvable point~\cite{Changlani_PRL}. 
While the \textit{classical} extensive degeneracy for the kagome and hyper-kagome lattice has a long history, 
the connection to the quantum case in the spin-1/2 $XXZ$ Hamiltonian, 
\begin{equation}
	H_{XXZ}[J_z] =  J \sum_{\langle i,j \rangle} S^{x}_{i} S^{x}_{j} + S^{y}_{i} S^{y}_{j} + 
	    J_{z}  \sum_{\langle i,j \rangle} S^{z}_{i} S^{z}_{j} 
\label{eq:XXZ}
\end{equation} 
at $H_\textrm{XXZ}[J_z=-1/2,J=1]$ (notated as $H_{XXZ0}$~\cite{Jaubert2016}), has not been entirely explored. 
$S_i$ are spin-1/2 operators on site $i$, $\langle i,j \rangle$ refer to nearest neighbor pairs 
and $J$ (set to 1 throughout the paper) and $J_z$ are the $XY$ and Ising couplings respectively. Ref.~\cite{Changlani_PRL} showed that the degeneracy exists in \emph{all} $S_z$ sectors 
and all finite (or infinite) system sizes. Numerical investigations on the highly symmetric $36d$ cluster 
showed how the $XXZ0$ point on the kagome lattice is embedded in the wider phase diagram. 

\begin{figure}
\centering
\resizebox{0.99\linewidth}{!}{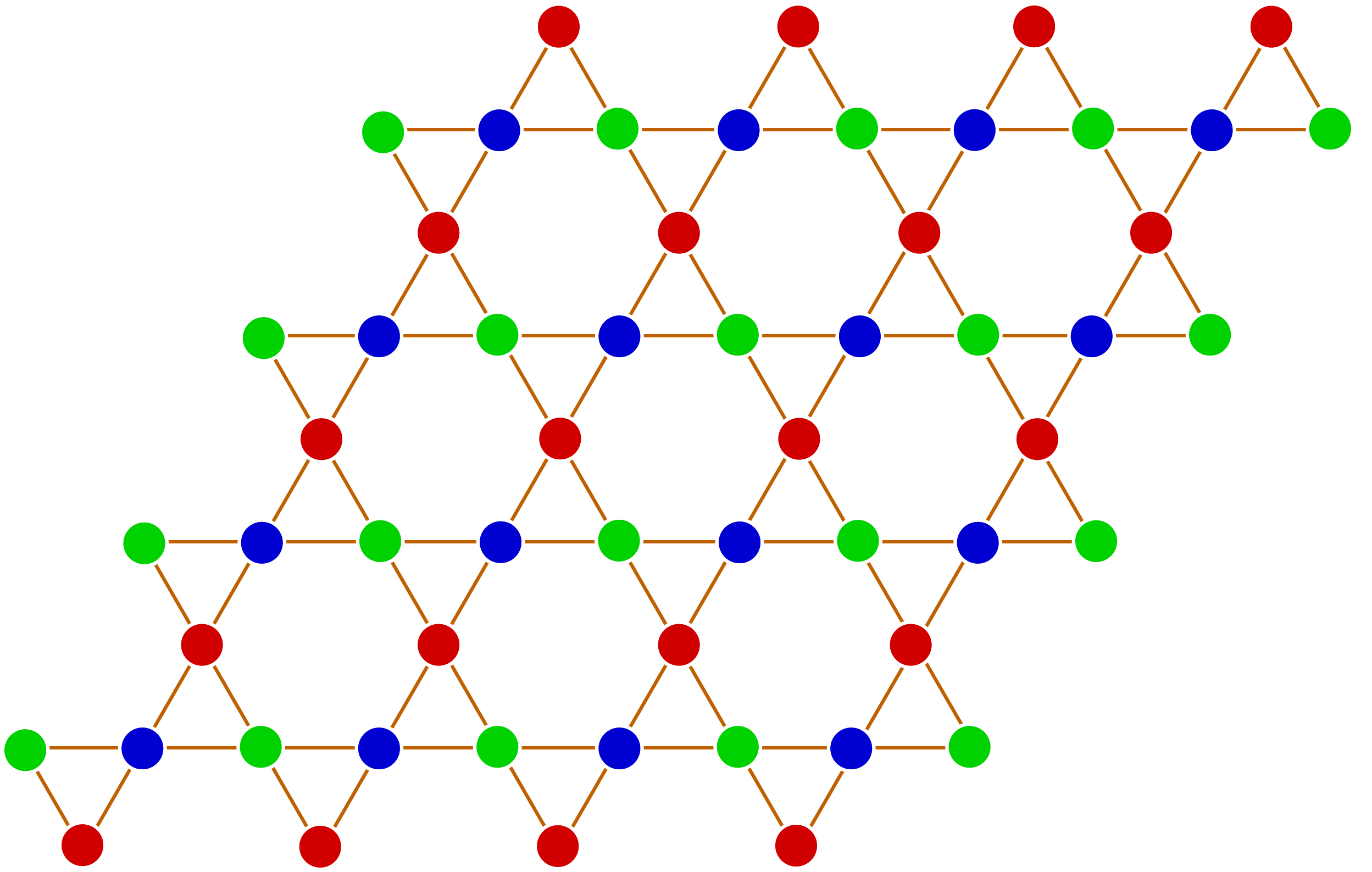} 
\\
\vspace{5mm}
\resizebox{0.99\linewidth}{!}{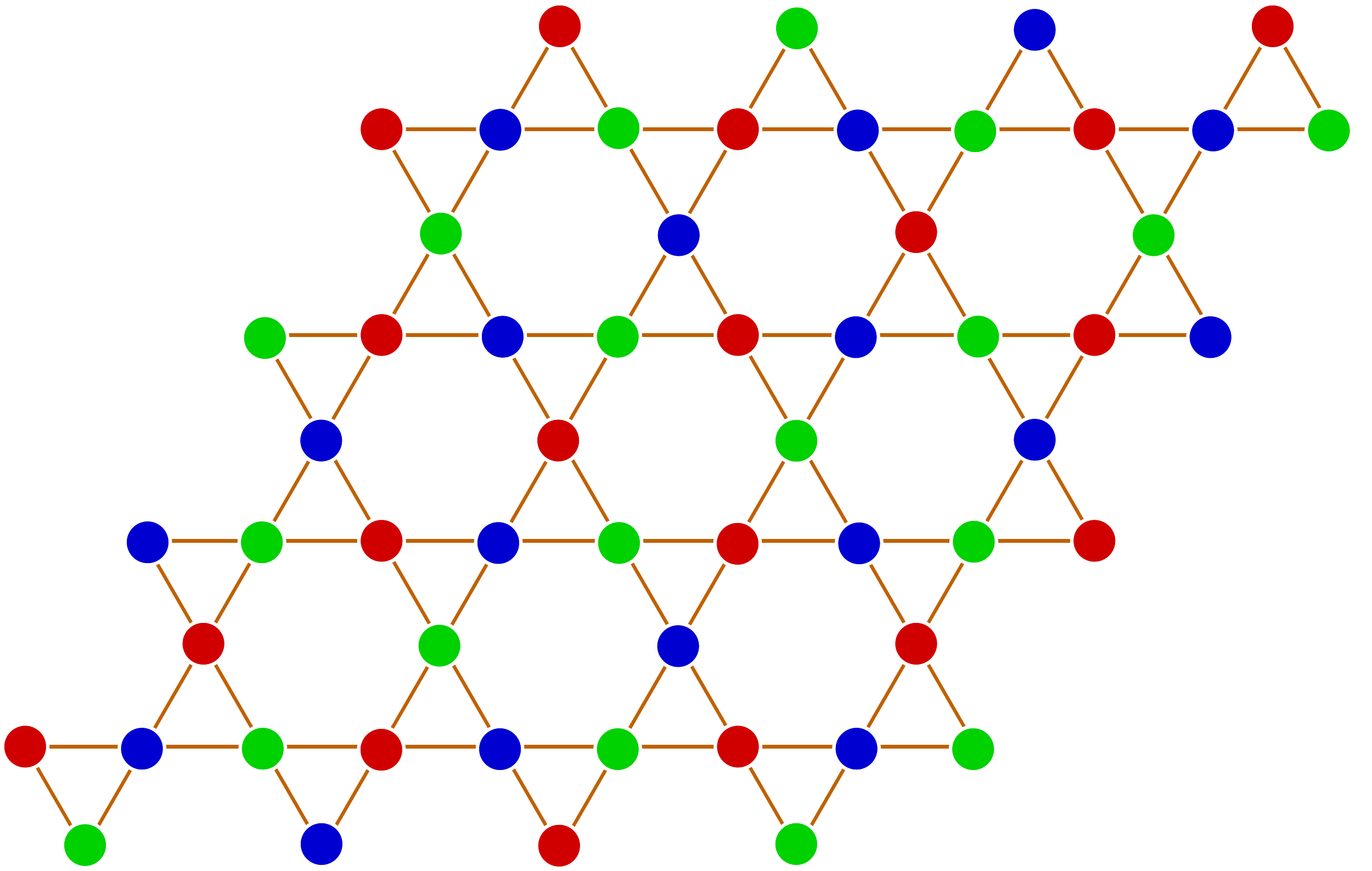} 
\caption{(Color online) Two representative three-colorings on the kagome lattice corresponding to the 
$q=0$ and $\sqrt{3} \times \sqrt{3}$ solutions. The colors red, blue and green represent the classical 
120$^\circ$ states or their quantum equivalents.
}
\label{fig:colorings} 
\end{figure}	

At $J_z=-1/2$, the exact solutions apply to any lattice of triangular motifs with the Hamiltonian of the form, 
\begin{equation} 
H = \sum_{\bigtriangleup} H_{XXZ0}(\bigtriangleup) \label{eq:triangleH}
\end{equation}
where $H_{XXZ0}(\bigtriangleup)$ is the $XXZ0$ Hamiltonian on a single triangle $\bigtriangleup$
as long as the vertices are consistently colorable by three colors such that 
no two vertices connected by a bond have the same color. For the kagome lattice, we show 
representative three-colorings in Fig.~\ref{fig:colorings} which depict the so-called $q=0$ and $\sqrt{3}\times\sqrt{3}$ patterns~\cite{Harris3color}. 
Other three-colorable lattices include the triangular lattice, 
the Shastry-Sutherland lattice the hyperkagome lattice, 
the squagome lattice and the icosidodecahedron. 

In this work, we employ the quantum three-colorings as a means of gaining analytic intuition for the physics near the highly degenerate $J_z=-1/2$ point. 
Our work will highlight the relevance of this point in controlling the physics seen in the Heisenberg regime i.e. $J_z=1$. 
For this purpose, we decompose the $XXZ$ Hamiltonian~\eqref{eq:XXZ} as,
\begin{eqnarray}
	H_{XXZ}[J_z] &=&  H_{XXZ0} + \Big( J_{z} + \frac{1}{2} \Big)  \sum_{\langle i,j \rangle} S^{z}_{i} S^{z}_{j} \\ 
	             &=&  H_{XXZ0} + \Big( J_{z} + \frac{1}{2} \Big)  H_{zz} 
\label{eq:XXZ0plusJz}
\end{eqnarray} 
and ask if it is possible to simultaneously minimize both parts of the Hamiltonian. 
While this is not possible in the most general circumstances, 
we find that at high magnetization (equivalently, low fillings in the hardcore bosonic language) 
the Hamiltonian is "frustration free" i.e. it is \textit{indeed} possible to achieve this minimization. 

Since the map between spin 1/2 and hardcore bosons is used often in the paper, we clarify the terminology associated with it. 
Down spins in a background of up spins are equivalent to hardcore bosons in a vacuum and thus we 
interchangeably use the words "filling" and "magnetization" in the course of our discussions. 
More precisely, the spin ($S_i$) and hardcore boson operators ($b_i$) are related as,
\begin{equation}
b_i^{\dagger}=S_i^{+} \;\; b_i = S_i^{-} \;\; n_i = b_i^{\dagger}b_i = \frac{1}{2} - S_i^{z}
\end{equation}
and thus the XXZ Hamiltonian reads,
\begin{equation}
H_{XXZ}(J_z) =  \frac{1}{2} \sum_{\langle i,j \rangle} b^\dagger_i b_j + \text{h.c.} + J_z \sum_{\langle i,j \rangle} n_{i} n_{j} + d 
\label{eq:HXXZ_rewrite0}
\end{equation}
where $d$ is a constant in a given magnetization sector that equals $J_z \left(
\frac{N}{2} - 2 \sum_{i} n_i \right)$ for a $N$ site kagome lattice.
We also use the term "magnon" to denote the wavefunction of one down spin in a sea of up spins, or equivalently 
the wavefunction of a single hardcore boson in vacuum. 

The remainder of the paper is organized as follows. In Sec.~\ref{sec:recap}, we recapitulate the 
nature of the exact (ground state) solutions for $J_z=-1/2$ and why they exist in every magnetization sector. 
For this we define quantum three-colors, the quantum version of the $120^{\circ}$ 
classical ground states, which provides a convenient choice of variables for explaining several of our numerical observations. 
In Sec.~\ref{sec:rcl_combined}, we develop the concept of resonating color loops (RCL) which is the basis of  
an exact mapping relating the coloring wavefunctions to magnons. We discuss in detail the crucial effects 
due to $S_z$ (or number) projection. Using the RCL construct, in Sec.~\ref{sec:rcl_flatband}, we revisit 
the more familiar localized and topological magnon modes, which arise from the flatband that exists 
on the kagome lattice. We show that each such mode has a direct connection to a RCL. In Sec.~\ref{sec:lowdensity}
these ideas are further extended to express exact many body ground state wavefunctions for special high magnetizations as projected 
quantum three-coloring wavefunctions. We find that for these special magnetization sectors, the exact ground state, a 
quantum three coloring superposition, holds for all $J_z\geq-1/2$ which shows the equivalence of the Ising, Heisenberg and $XY$ regimes. 

For the case of zero magnetization, we have investigated the relevance of the $J_z=-1/2$ point (and hence the three-coloring manifold) 
by performing large scale density matrix renormalization group (DMRG) calculations for a large range of $J_z$ in Sec.~\ref{sec:dmrg}. These results extend 
the results of previous exact diagonalizations~\cite{Changlani_PRL} to bigger systems which support $H_{XXZ0}$ being 
a quantum critical point in the $XXZ$ phase diagram. In Sec.~\ref{sec:discuss}, we conclude by summarizing our results and suggesting future avenues for further exploration. 

\section{Quantum three-colors and the exact solution of $H_{XXZ0}$} 
\label{sec:recap}
We state the central result of Ref.~\cite{Changlani_PRL}, where it was proved that any Hamiltonian of the form 
of Eq.~\eqref{eq:triangleH} for $J_z=-1/2$, has ground states of the form, 
\begin{equation}
	| C \rangle \equiv  P_{S_z} \Big( \prod_{\text{valid}} \otimes | \gamma_s \rangle \Big)
\label{eqn:GroundState}
\end{equation}
where $\{ |\gamma_s\rangle = |r\rangle,|b \rangle \text{ or } |g\rangle \}$, denoted as colors on site $s$ are defined as,
\begin{eqnarray} 
|r \rangle &\equiv& \frac{1}{\sqrt{2}} \Big(|\uparrow \rangle + |\downarrow \rangle \Big) \nonumber \\
|b \rangle &\equiv& \frac{1}{\sqrt{2}} \Big(|\uparrow \rangle + \omega |\downarrow \rangle \Big) \nonumber \\
|g \rangle &\equiv& \frac{1}{\sqrt{2}} \Big(|\uparrow \rangle + \omega^2 |\downarrow \rangle \Big)
\end{eqnarray}
where $\omega=e^{i2\pi/3}$. Taking the quantization axis to be the $z$-axis, 
the colors correspond to spin directions in the $x-y$ plane that are at $120^{\circ}$ 
relative to one another. Valid colorings satisfy the three-coloring condition i.e. exactly one $|r\rangle$, one $|b\rangle$ and one $|g\rangle$ per triangular motif. 
These are depicted by colors red, blue and green respectively in our figures. 
$P_{S_z}$ projects into a particular total $S_z$ sector.   

The construction~\eqref{eqn:GroundState} is loosely referred to as the three-coloring condition and any such many body state 
which satisfies the constraint conditions is a three-coloring state. Such states have primarily been studied in the context of the classical kagome 
antiferromagnet at the Heisenberg point~\cite{Harris3color,Henley3color,Huse_Rutenberg,Chalker3color,Sachdev92, vonDelft_Henley, 
Cepas, Jaubert2016,Castelnovo_three_coloring}. 

Classically, a Luttinger-Tisza analysis~\cite{Luttinger_Tisza} of $H_{XXZ}$ shows that 
$J_z=-1/2$ is a critical point in the kagome phase diagram. To see this, 
we recast Eq.~\eqref{eq:XXZ} in reciprocal space,
\begin{align}
\sum_\mathbf{q} \bigg(
\tilde{\mathbf{S}}_{XY}(\mathbf{q})^T & 
\cdot \left[ \tilde{J}(\mathbf{q}) \right] \cdot
 \tilde{\mathbf{S}}_{XY}(\mathbf{-q}) \nonumber \\
+ &
J_z \; \tilde{\mathbf{S}}_Z(\mathbf{q})^T \cdot \left[ \tilde{J}(\mathbf{q}) \right] \cdot
\tilde{\mathbf{S}}_Z(\mathbf{-q})
\bigg)
\label{eq:reciprocal}
\end{align}
where,
\begin{align}
\tilde{\mathbf{S}}_{XY}(\mathbf{q}) & = \frac{1}{L_X L_Y} \sum_{\mathbf{r}}
e^{- i \mathbf{q} \cdot \mathbf{r}} \left( 
\begin{matrix}
\mathbf{S}^{XY}_{\mathbf{r},1} & \mathbf{S}^{XY}_{\mathbf{r},2} & \mathbf{S}^{XY}_{\mathbf{r},3}
\end{matrix}
\right)^T \\
\tilde{\mathbf{S}}_{Z}(\mathbf{q}) & = \frac{1}{L_X L_Y} \sum_{\mathbf{r}}
e^{- i \mathbf{q} \cdot \mathbf{r}} \left( 
\begin{matrix}
\mathbf{S}^z_{\mathbf{r},1} & \mathbf{S}^z_{\mathbf{r},2} & \mathbf{S}^z_{\mathbf{r},3}
\end{matrix}
\right)^T \\
\left[ \tilde{J}(\mathbf{q}) \right] & =
\frac{1}{2} \left(
\begin{matrix}
0 & 1 + e^{i \mathbf{q} \cdot \mathbf{a_2}} & 1 + e^{i \mathbf{q} \cdot 
(\mathbf{a_2} - \mathbf{a_1} )} \\
1 + e^{-i \mathbf{q} \cdot \mathbf{a_2}} & 0 & 1 + e^{-i \mathbf{q} \cdot \mathbf{a_1}} \\
1 + e^{-i \mathbf{q} \cdot 
(\mathbf{a_2} - \mathbf{a_1} )} 
& 1 + e^{i \mathbf{q} \cdot \mathbf{a_1}} & 0
\end{matrix}
\right)
\label{eq:LT}
\end{align}
$\mathbf{a_1}$, $\mathbf{a_2}$ are the primitive lattice vectors (considering one up-triangle with three sites as the Kagome unit cell),
and $\mathbf{q}$ is restricted to the first Brillouin zone. 
$\mathbf{S}_{\mathbf{r,\mu}}$ is a classical spin of unit magnitude
at site $\mathbf{r},\mu$. $\mathbf{S}^{XY}_{\mathbf{r},\mu}$ and $\mathbf{S}^{Z}_{\mathbf{r},\mu}$
are the projections of the unit vector $\mathbf{S}_{\mathbf{r,\mu}}$ 
on to the $x-y$ plane and $z$ axis respectively.

For the classical ground state, the two terms in Eq.~\eqref{eq:reciprocal} 
are competing. For $J_z < -1/2$, the second term in Eq.~\eqref{eq:reciprocal} 
wins giving a unique ferromagnetic ground state at $\mathbf{q}=0$ with all spins 
pointing in the $z$ direction. For $J_z > -1/2$,  the first 
term in Eq.~\eqref{eq:reciprocal} wins, giving a flat-band solution with all 
$\mathbf{q}$ being classically degenerate to each other in energy, 
with the spins oriented in the $x-y$ plane. They give rise to an extensively degenerate 
ground state manifold since there are infinite ways
 in which these classically degenerate solutions at different $\mathbf{q}$ may be linearly combined 
while respecting the Luttinger-Tisza condition $\sum_\mathbf{q} 
\tilde{\mathbf{S}}_{XY}(\mathbf{q}) \cdot \tilde{\mathbf{S}}_{XY}(\mathbf{-q}) = 1$.
At the Heisenberg point, 
this extensively degenerate classical ground state manifold has also been noted
in the literature before~\cite{Elser_89,Lhuillier_05,Wen_12}.

Since the classical spins lie in the $x-y$ plane for $J_z > -1/2$, they are impervious to the $J_z$ term.
Any state that then locally satisfies the three-coloring 
($120^{\circ}$) condition is a classical ground state for $J_z\geq-1/2$ and $J_z<1$ (there is an additional classical phase transition at the Heisenberg point 
$J_z=1$, which we do not explore). Then, exactly at the $J_z=-1/2$ point, the classical and quantum solutions develop 
a one to one correspondence. However, an important difference from the classical solution, is that quantum mechanically, the wavefunctions must have definite 
total $S_z$ as the $XXZ$ Hamiltonian has $U(1)$ symmetry. Therefore, projecting each three-coloring solution to each $S_z$ sector must \emph{also} be an exact ground state, 
thereby justifying the projection in Eq.~\eqref{eqn:GroundState}. Conversely, this also implies that this exactly solvable point exists in all $S_z$ sectors. 

\begin{figure}
\centering 
\includegraphics[width=\linewidth]{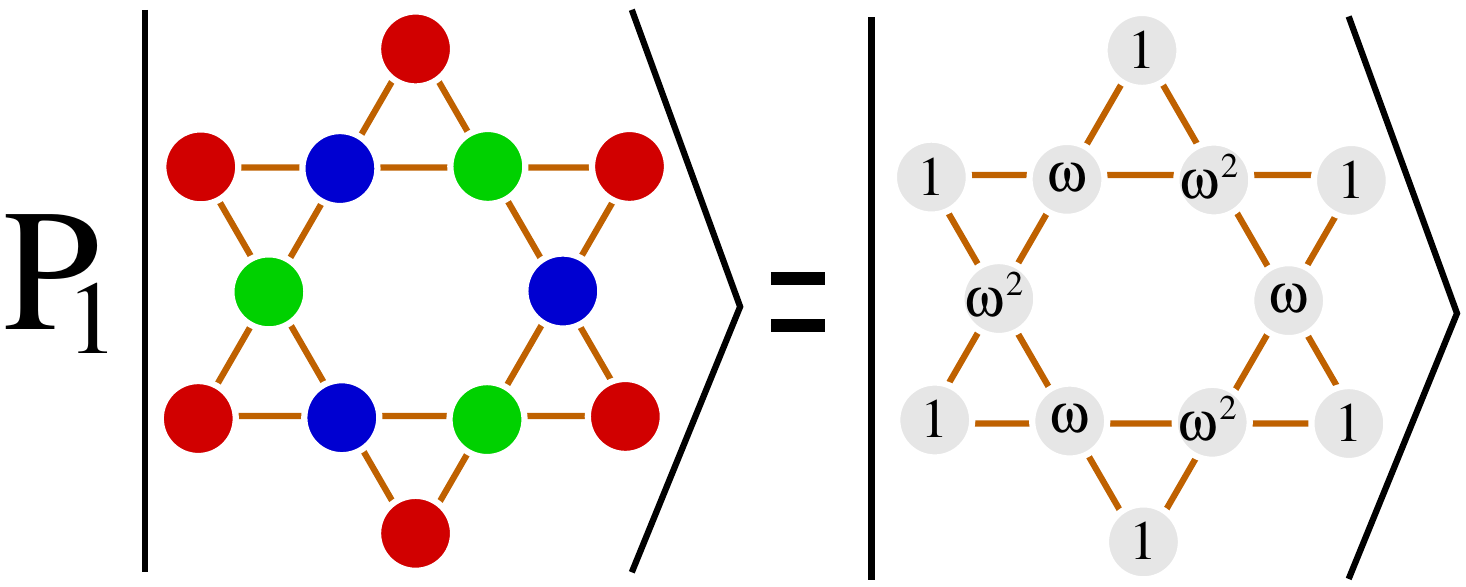}
\caption{(Color online) Representative example of a single magnon state with amplitudes 
$1,\omega,\omega^2$ in the three coloring basis, written as a many-body coloring wavefunction with a projection operator.}
\label{fig:color_magnon_map} 
\end{figure}

The three coloring wavefunctions when projected to the one particle sector (or one spin-down sector), 
can be viewed as the wavefunction of a single particle on the kagome lattice. 
One such example has been represented in Fig.~\ref{fig:color_magnon_map}. Depending on the color associated with the site, 
the amplitudes are $1$, $\omega$ or $\omega^2$. Taking linear combinations of single particle wavefunctions (i.e. adding their amplitudes site by site) 
is exactly equivalent to taking linear combinations of projected colorings, since $P_1 |C_1\rangle - P_1 |C_2 \rangle = P_1 (|C_1\rangle - |C_2 \rangle)$. 
This concept will be used in the next section when discussing resonating color loops.

\begin{figure}
\centering
\includegraphics[width=\linewidth]{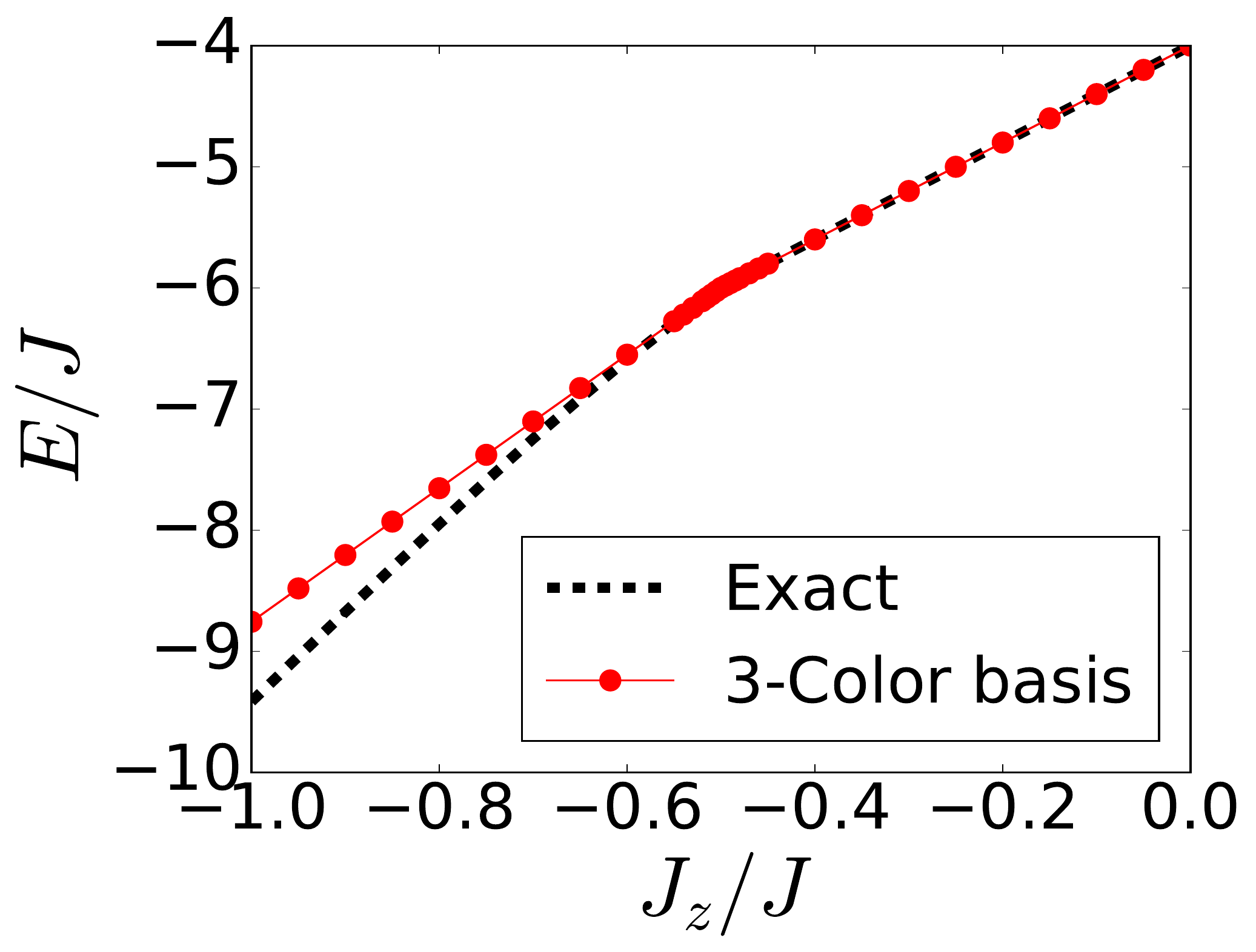}
\includegraphics[width=\linewidth]{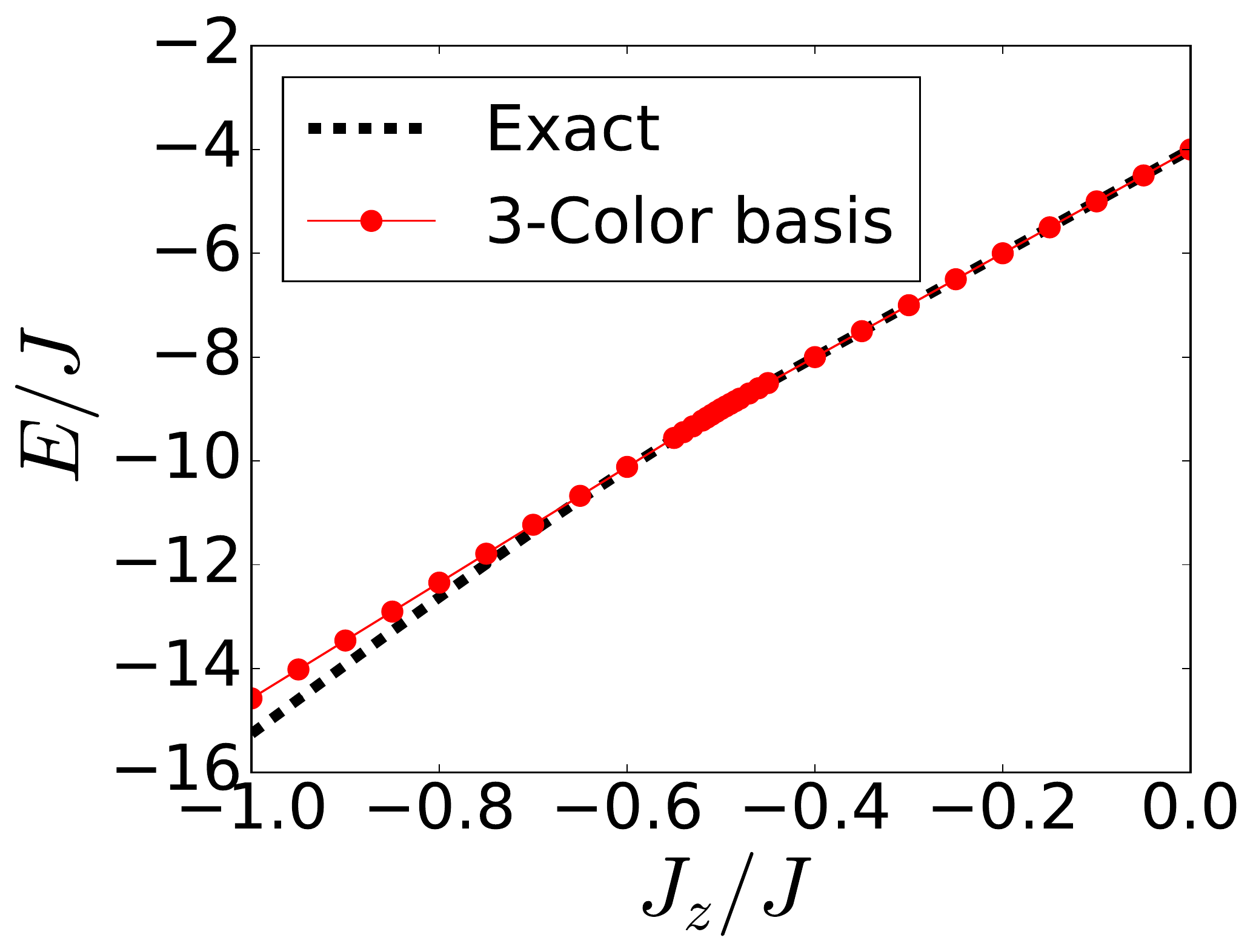}
\caption{Comparison of ground state energies from exact diagonalization 
and diagonalization in the three-color basis as a function of $J_z$ ($J=1$) 
for the (top panel) $4\times2\times3$ torus for $S_z=8$ (1/6 filling of bosons) and (bottom panel) 
36d cluster for $S_z=14$ (1/9 filling of bosons), in the range $-1 \leq J_z \leq 0$. 
Since the Hamiltonian is frustration free for $J_z \geq -1/2$, the exact ground state 
solution hold for arbitrary $J_z \geq -1/2$.} 
\label{fig:energies_4x2x3} 
\end{figure}	

The total number of three-coloring ground states scales exponentially with system size. However, 
there are two subtleties to be considered when counting the exact number of linearly independent solutions 
when projecting to definite $S_z$. First, when one interchanges colors (consistently for all sites), the new coloring $|C'\rangle$ is not linearly independent of the original 
one $|C\rangle$. This can be seen by redefining,
\begin{equation}
	|\downarrow \rangle' \equiv \omega |\downarrow \rangle
\end{equation} 
which is equivalent to the transformation (from old to new variables)
\begin{eqnarray}
r \rightarrow g  \\ 
b \rightarrow r  \nonumber \\ 
g \rightarrow b \nonumber
\end{eqnarray} 
Under this transformation each spin configuration (and hence the overall wavefunction) 
is only rescaled by a constant phase $\omega^{N_{\downarrow}}$ where $N_{\downarrow}$ is the number of down spins. 
A similar transformation holds for $|\downarrow\rangle' \equiv \omega^2 |\downarrow \rangle$ 
which leads to $r \rightarrow b$, $b \rightarrow g$, $g \rightarrow r$. Thus, 
these three-colorings are not linearly independent and should not be counted more than once.

The second subtlety when counting the number of colorings is that not all colorings remain linearly independent 
when projected to definite total $S_z$. This is best exemplified by considering the case of 
the fully ferromagnetic sector. Here, even though the number of three colorings is exponential, 
there is only one unique solution possible. Thus, to determine the precise number of linearly independent 
many body states, we evaluate the rank ($R(S)$) of the overlap matrix $S_{CC'}=\langle C | C'\rangle$. 
The matrix elements are calculated efficiently and the matrix numerically diagonalized for this purpose. 
(Details of the calculation of the matrix elements in this non-orthogonal basis 
have been discussed at length in the supplemental information of Ref.~\cite{Changlani_PRL} and are 
hence not presented here.) This enumeration of three-coloring states and their counting 
is an essential part of the diagonalizations we perform in the restricted subspace of the full Hilbert space. 

Till this stage, our discussion has focused on the $J_z=-1/2$ point, which is only one point in the parameter space of the $XXZ$ model. 
However, as mentioned in the introduction, the concept of color degrees of freedom 
and three-coloring states is useful more generally; we will show this more explicitly in the subsequent sections. 
For example, in a quest to minimize both parts of Eq.~\eqref{eq:XXZ0plusJz}, we have diagonalized the $XXZ$ Hamiltonian in 
the three-coloring basis numerically by solving the generalized eigenproblem,
\begin{equation}
	{\bf H x} = E {\bf S x}
\end{equation}
where ${\mathbf H_{CC'}}=\langle C|H|C'\rangle$, $E$ is the eigen energy and 
$\mathbf{x}$ is the eigenvector of coefficients of three-color basis states. 
The ground state energy is compared to the exact ground state energy in the 
full (Ising) basis. Fig.~\ref{fig:energies_4x2x3} shows the results of these comparisons 
for the $4\times2\times3$ cluster at $2/3$ magnetization ($1/6$ filling) and the $36d$ 
cluster at $7/9$ magnetization ($1/9$ filling).

The three-coloring states (consisting only of colorings satisfying the three-coloring constraint) 
does not form a complete set in a specified $S_z$ sector - i.e. it can not describe an arbitrary wavefunction. 
However, for the representative examples demonstrated here we \textit{do} obtain the \textit{exact} energy 
for $J_z\geq-1/2$. Thus, for these cases the exact wavefunctions lie completely in the three-coloring manifold with a total ground state 
energy equal to $E_{XXZ0} + (J_z+1/2) \left(\frac{N}{2} - 2 \sum_{i} n_i \right)$ for a $N$ site kagome lattice. 
These numerical findings suggest the existence of an analytic way of understanding the three-coloring superposition and we 
will develop the appropriate concepts for proving that this is indeed the case.

\section{Resonating color loops}
\label{sec:rcl_combined}
In this section we will develop the machinery to generate, on some lattices and at low density,  simultaneous ground states of $H_{XXZ0}$ and $H_{zz}$ 
making them frustration free ground states of $H_{XXZ0}+ (J_z + \frac{1}{2}) H_{zz}$. Unfortunately, no single three-coloring is such a ground state, 
except in the extreme case of a fully polarized state. 
Instead, we need to construct linear combinations of three-colorings; such states are already ground states of $H_{XXZ0}$ 
and so our focus will be developing linear combinations which minimize $H_{zz}$ at low density.  

\begin{figure*}
\centering
\includegraphics[width=0.8\linewidth]{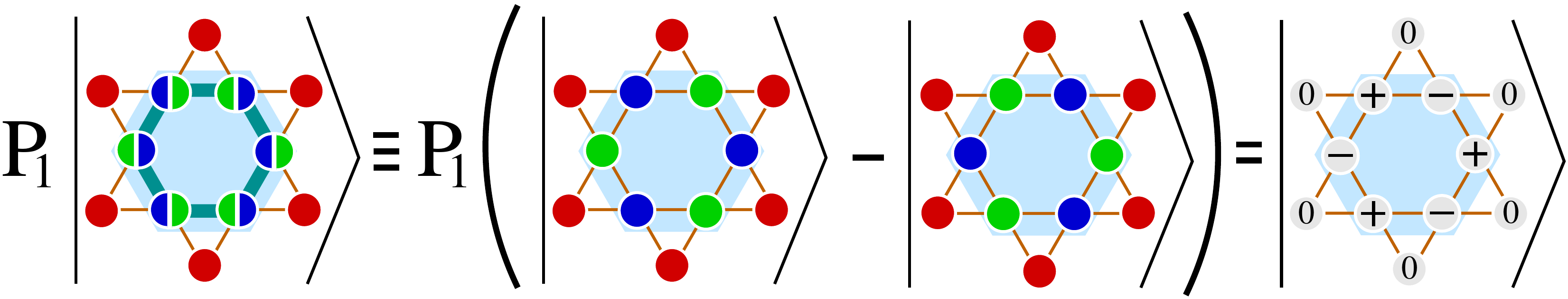}
\includegraphics[width=0.8\linewidth]{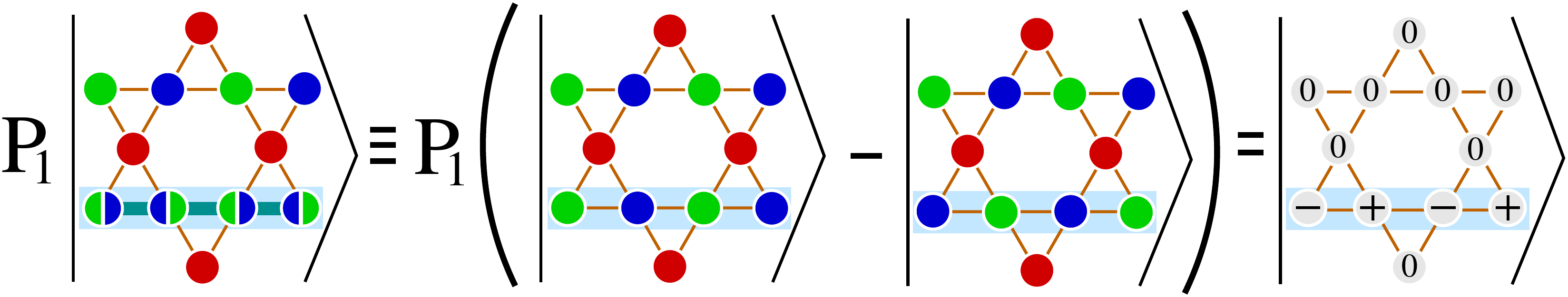}
\caption{(Color online) Definition of resonating color loops on a kagome lattice. 
Each RCL is obtained by taking a difference of two three-colorings, which differ only on a single two-color loop. 
In the top panel, the RCL is located on a hexagon and in the bottom panel it is located on a topological (non-contractible) 
loop, here winding along the x direction. The RCLs when projected to a single spin-down (magnon) sector 
are exactly equal to localized or topological magnons on the kagome lattice up to a
(projective) phase, and an innocuous normalization.}
\label{fig:color_map_david_star} 
\end{figure*}

The key tool in accomplishing this task will be resonating color loops (RCL).  A RCL is generated by taking a single 
closed ``two-color" loop (comprising, say of green and blue colors) and replacing it with a linear combination of the two different green-blue colorings over that loop 
with a relative minus sign between them. For example, consider the closed loop corresponding to the hexagonal plaquette on the kagome lattice. Then, 
the quantum state 
\begin{equation}
|\textrm{RCL}\rangle = |gbgbgb\rangle  - |bgbgbg\rangle
\label{eq:rcl_eqn} 
\end{equation}
is what we define as a green-blue RCL (see Fig.~\ref{fig:color_map_david_star}). 
For the purpose of this work, the definition of the RCL adopted is always of the form Eq.~\eqref{eq:rcl_eqn}. However, 
in principle, it is possible to generalize the concept of RCL to other linear superpositions, with certain desirable properties.
The local resonating structure of RCLs is thus reminiscent of resonating valence bond (RVB) states.  

Consider a fixed three-coloring with some number of two-color loops which are adjacent only to a third color, i.e. an \textit{isolated} two-color loop (ICL). 
Any $k$ ICL can be replaced with $k$ RCL and the resulting state will be a linear superposition of $2^k$ three-colorings. This follows because if an entire 
two-colored loop (say of green and blue) is surrounded by red, then swapping green and blue within that loop still leaves no edge with the same color on both vertices. 
As an example, consider the $\sqrt{3} \times \sqrt{3}$ coloring of the kagome lattice~(Fig.~\ref{fig:colorings}). 
This coloring has isolated two-color hexagonal loops.  We can take any number of these hexagons and turn them into RCLs. 
Alternatively, on the $q=0$ coloring on the kagome lattice~(Fig.~\ref{fig:colorings}) 
there are isolated non-contractible loops which can be turned into a RCL. It is interesting to note that on a coordination-4 lattice of triangles 
every site is part of an isolated two-color loop.  

\begin{figure}
\centering 
\includegraphics[width=\linewidth]{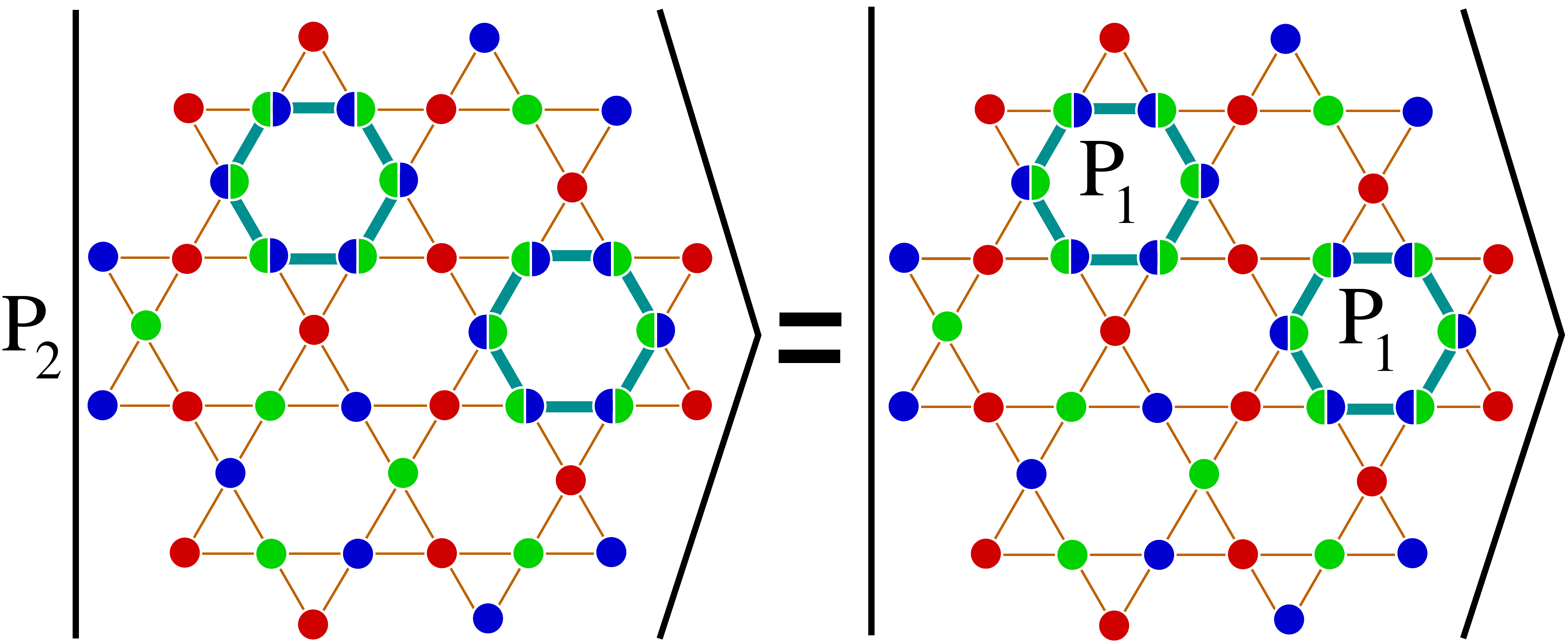}
\caption{(Color online)
A representative example of projection on to the $k = 2$ spin-down sector on a 
configuration with two RCLs. The projection properties of RCLs ensure 
localisation of bosons/spin downs to localized hexagons.
\label{fig:keq2_proj_eg}
}
\end{figure}

Now that we have a linear combination of three-colorings generated by replacing ICL with RCL, we can consider the role of projection on these states.  
In particular, we will see that if we globally project a state with $k$ RCL into the sector of $k$ spin-down (i.e. $P_k$), then  there will be \emph{exactly one} 
spin-down constrained to each RCL and \emph{no} spin-down outside the RCL.  A $k=2$ representative
example is shown in Fig. \ref{fig:keq2_proj_eg}. 

To see this why this particle localization happens,
we first note that
the difference of two colorings is destroyed by projecting into the fully spin-down sectors
on a given RCL, i.e. 
\begin{equation}
P^{\text{RCL}}_0 \left( |C^{1}_m \rangle - |C^{2}_m \rangle\right) = 0
\label{eq:RCL_localize}
\end{equation} 
where $C^{1}_m$ and $C^{2}_m$ are arbitrary colorings on the motif denoted by $m$. 
It then follows that  $P_0|\textrm{RCL}\rangle=0$ (here it is important the RCL is the difference of two loops).  
Now, let us  consider, as an example, $P_2$ applied to a quantum state with $2$ RCL and decompose $P_{2}$  into the sum of tensor products of projectors over the two RCLs and the rest of the system respectively, written explicitly
as
\begin{align}
P_2 = & P^{\text{RCL}_1}_2 \otimes  P^{\text{RCL}_2}_0 \otimes P^{\text{rest}}_0 + \nonumber \\
& P^{\text{RCL}_1}_0 \otimes  P^{\text{RCL}_2}_2 \otimes P^{\text{rest}}_0 + \nonumber \\
& P^{\text{RCL}_1}_0 \otimes  P^{\text{RCL}_2}_0 \otimes P^{\text{rest}}_2 + \nonumber \\
& P^{\text{RCL}_1}_1 \otimes  P^{\text{RCL}_2}_1 \otimes P^{\text{rest}}_0 + \nonumber \\
& P^{\text{RCL}_1}_1 \otimes  P^{\text{RCL}_2}_0 \otimes P^{\text{rest}}_1 + \nonumber \\
& P^{\text{RCL}_1}_0 \otimes  P^{\text{RCL}_2}_1 \otimes P^{\text{rest}}_1 \nonumber \\
\implies P_2 = & P^{\text{RCL}_1}_1 \otimes  P^{\text{RCL}_2}_1 \otimes P^{\text{rest}}_0 
\end{align}
The last equality follows from Eq.~\eqref{eq:RCL_localize}, as any term in the sum with $P^{\text{RCL}}_0$ on an RCL is destroyed.
This is schematically shown in Fig. \ref{fig:keq2_proj_eg}. The above generalizes
straightforwardly to $k$ RCLs projected to $k$ spin-down sector. Using this machinery, we thus 
have an ability to localize down-spins onto any ICL. This ability will allow us to minimize $H_{zz}$ by ensuring that two spin-downs are never nearest neighbors. 

In the next two sections we will see (1) that this argument (RCL when projected into a single spin-down) is essentially a quantum coloring language to describe kagome flatband magnons and (2) that on a variety of coordination-4 lattice of triangles such as the kagome lattice, the kagome ladder and the squagome lattice, at high magnetizations (i.e. 
at low fillings), the many body ground state is a tensor product of RCLs projected to that $S_z$ sector.

\section{Kagome flat band modes from resonating color loops} 
\label{sec:rcl_flatband}
In the previous section we considered how an RCL can be used to localize down spins on certain motifs. 
In this section we are going to consider systems with a single RCL being projected into the single spin-down sector (i.e. $P_1$) finding an 
exact correspondence between these projected RCLs and the localized and topological magnons~\cite{Richter_magnons} associated with the flatband of the kagome lattice.  
To understand this result, we first review the results of Ref.~\cite{Bergman_Balents} which explained the existence of kagome lattice flat band modes 
using a localized basis of single-particle orbitals. 

\begin{figure}
\centering
\includegraphics[width=1.\linewidth]{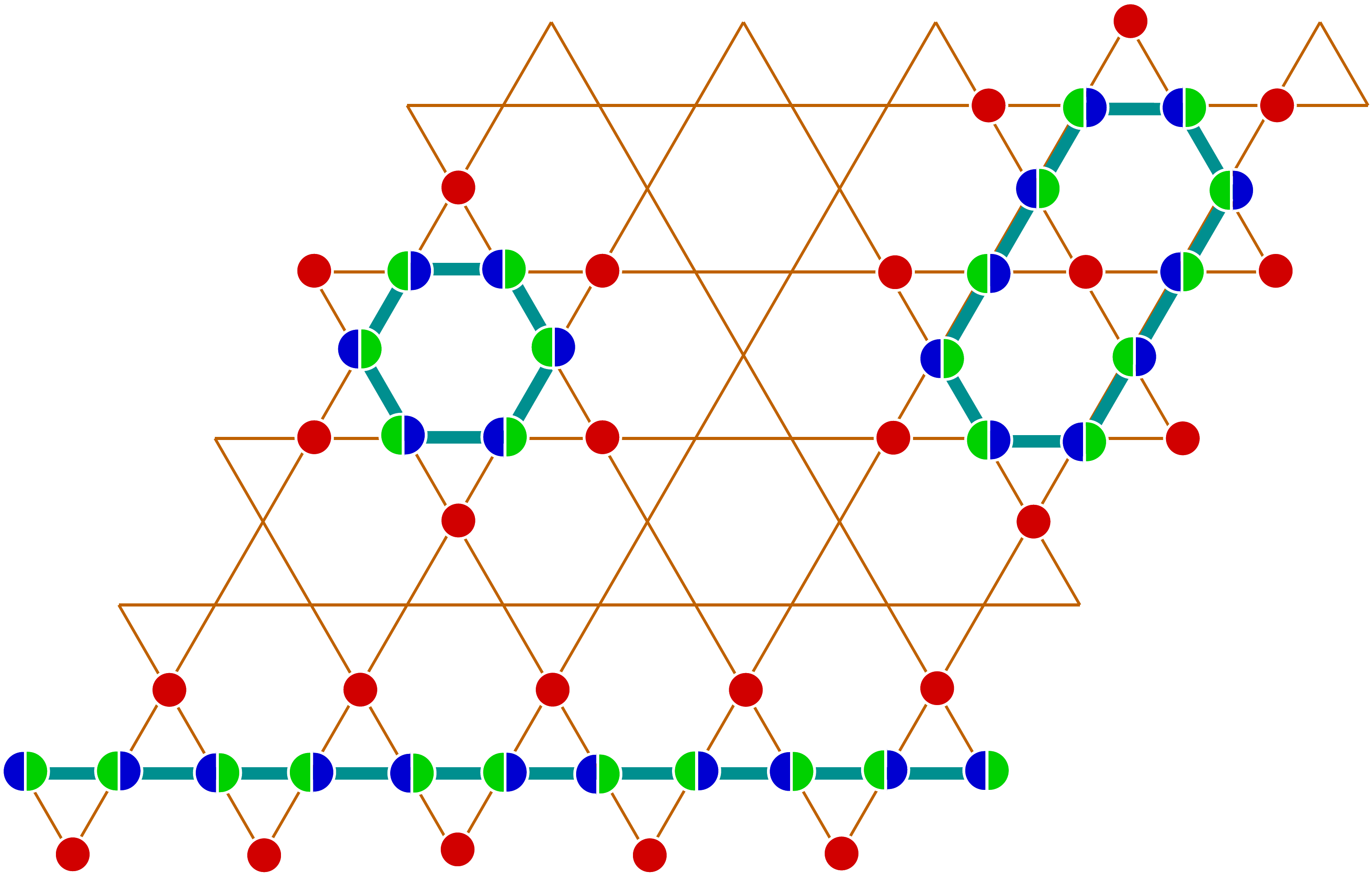}
\caption{(Color online)
Representative locations of localized and topological single particle modes
as resonating color loops are shown, including a 10 site loop that may
be thought of as a composition of two hexagonal localized modes.
Fig.~\ref{fig:color_map_david_star} shows how to transcribe the above RCL representation
into the magnon modes.
Apart from the \emph{single} RCL at a chosen representative location, the
rest of lattice is the same valid three-coloring, which makes the cancelation at all other sites 
exact.}
\label{fig:flatband_modes} 
\end{figure}	

First we note that the $XXZ$ Hamiltonian with a single down spin corresponds to the non-interacting tight binding model on the kagome lattice 
giving  three bands with dispersions, 
\begin{subequations}
\begin{eqnarray}
	\epsilon_0({\bf q}) &=& -t \\
	\epsilon_{\pm}({\bf q}) &=& \frac{t}{2} \Big(1 \pm \sqrt{3+2\Lambda({\bf q})}\Big)
\end{eqnarray}
\end{subequations}
where $\Lambda(\bf{q})= cos({\bf q \cdot a_1}) + cos (\bf{q \cdot a_2}) + cos (\bf{q \cdot a_3})$. 
For $t>0$, the flat band becomes the lowest
energy band and at ${\bf q}=0$, $\epsilon_{-}$ touches the flat band. 
Thus, on a kagome lattice on a finite torus (periodic boundary conditions), 
with $N$ unit cells with a finite momentum grid with $N$ points, there are $N+1$ single particle states at $\epsilon=-t$. 

The flatness of the band allows us to take linear combinations of single particle states freely 
while remaining eigenstates.  This leads to a useful and insightful representation 
that results in localized eigenstates~\cite{Bergman_Balents}, 
given by,
\begin{equation}
	{A_{R}}^{\dagger}=\frac{1}{\sqrt{L}} \sum_{j=1}^{L} (-1)^{j} b_{j}^{\dagger}
\label{eq:local_mode}
\end{equation} 
where $L$ is the length of any contractible loop of length $4m+2$ or non contractible loop of length $2m$ where $m$ is an integer 
and $j$ refers to the index of lattice sites numbered in a contiguous order. 
When $L=6$, this mode is localized on a hexagonal motif, and is represented in the right most side of the top panel 
of Fig.~\ref{fig:color_map_david_star} ignoring the normalization of $\sqrt{6}$. Intuitively, this mode can be understood 
using a simple quantum interference argument. The topology of the kagome is such that the "$+$" and "$-$" contributions from hopping on to the vertices pointing away from the hexagon 
cancel out destructively and thus such a localized state becomes an exact ground state of the tight binding Hamiltonian. 

We now identify the relation between the quantum coloring language and these localized single particle orbitals.  
By taking a single projected RCL on a hexagon shown in Fig.~\ref{fig:color_map_david_star} a pattern of alternating $(\omega-\omega^2)$ and $(\omega^2 - \omega)$ 
is obtained on the hexagon with 0 everywhere else. Up to an overall phase factor, 
the mode is identical to the alternating pattern of $+$ and $-$ described above. 
In fact, this argument holds for arbitrary $L=4m+2$, such as the 10 site loop (which can be alternately viewed as a superposition of 
two localized single-particle hexagon wavefunctions) which corresponds to a projected 10 site RCL (see Fig.~\ref{fig:flatband_modes}). 
Thus projected RCLs have the form as in Eq.~\eqref{eq:local_mode}.

The set of $N$ hexagon single particle modes is not completely linearly independent; 
the wavefunction of the $N^{th}$ hexagonal mode can be rewritten as a linear combination of the remaining $N-1$ modes~\cite{Bergman_Balents}.
Since the expected count of the lowest degenerate states is $N+1$, this leaves us with two modes to be determined. 
Ref.~\cite{Bergman_Balents} showed that these correspond to 
two topological modes, coming from any choice of two non-contractible loops along the two periodic directions on the torus. 
An example of such a loop in the horizontal direction is shown in Fig.~\ref{fig:flatband_modes} (bottom).
Once again, this topological magnon has a natural meaning in the basis of three colors, and as is shown in Fig.~\ref{fig:color_map_david_star}, 
it is identical to an RCL defined on a two-color loop along the horizontal direction.

We have thus shown that every single particle magnon corresponding to the kagome flat band is exactly an RCL of a certain type. 
This is particularly useful, because it allows us to freely swap concepts between two distinct languages. 
In particular, in the next section we will provide a new interpretation of many-body wavefunctions constructed at low magnon fillings, 
in terms of RCLs.

\section{Low density exact solutions from resonating color loops} 
\label{sec:lowdensity}

\begin{figure}
\centering
\includegraphics[width=0.65\linewidth]{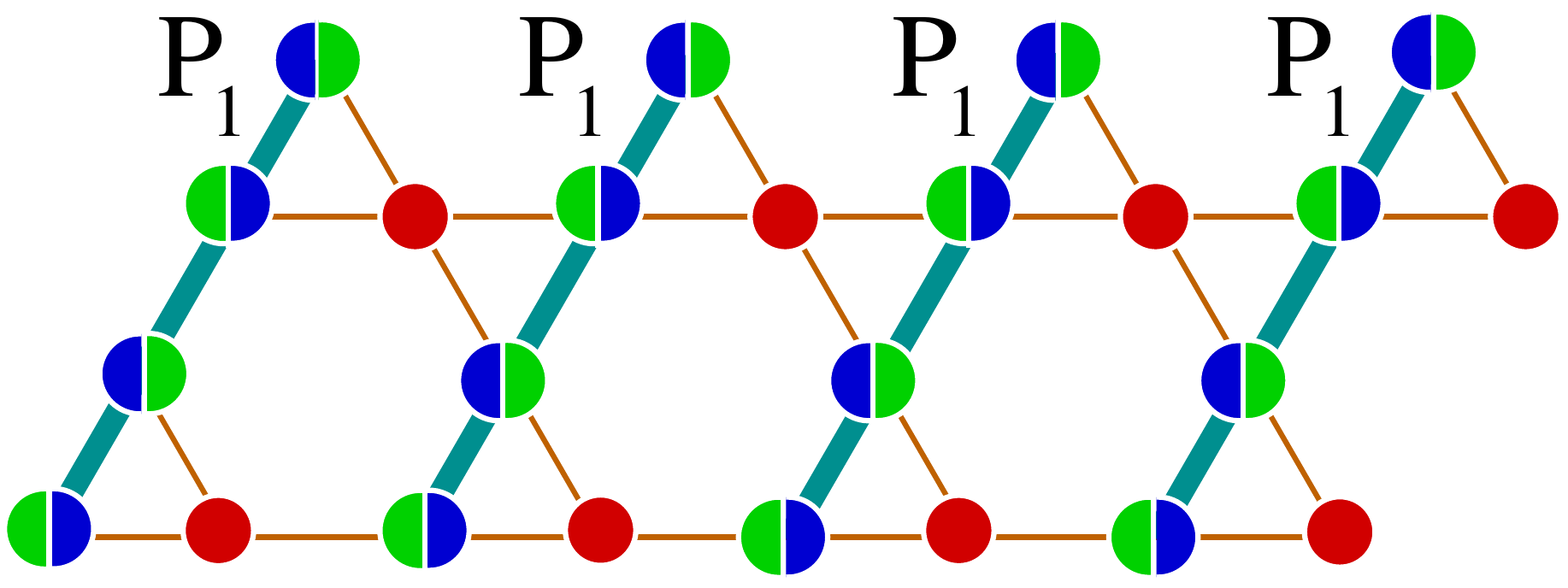}
\vspace{5mm} \\
\includegraphics[width=0.75\linewidth]{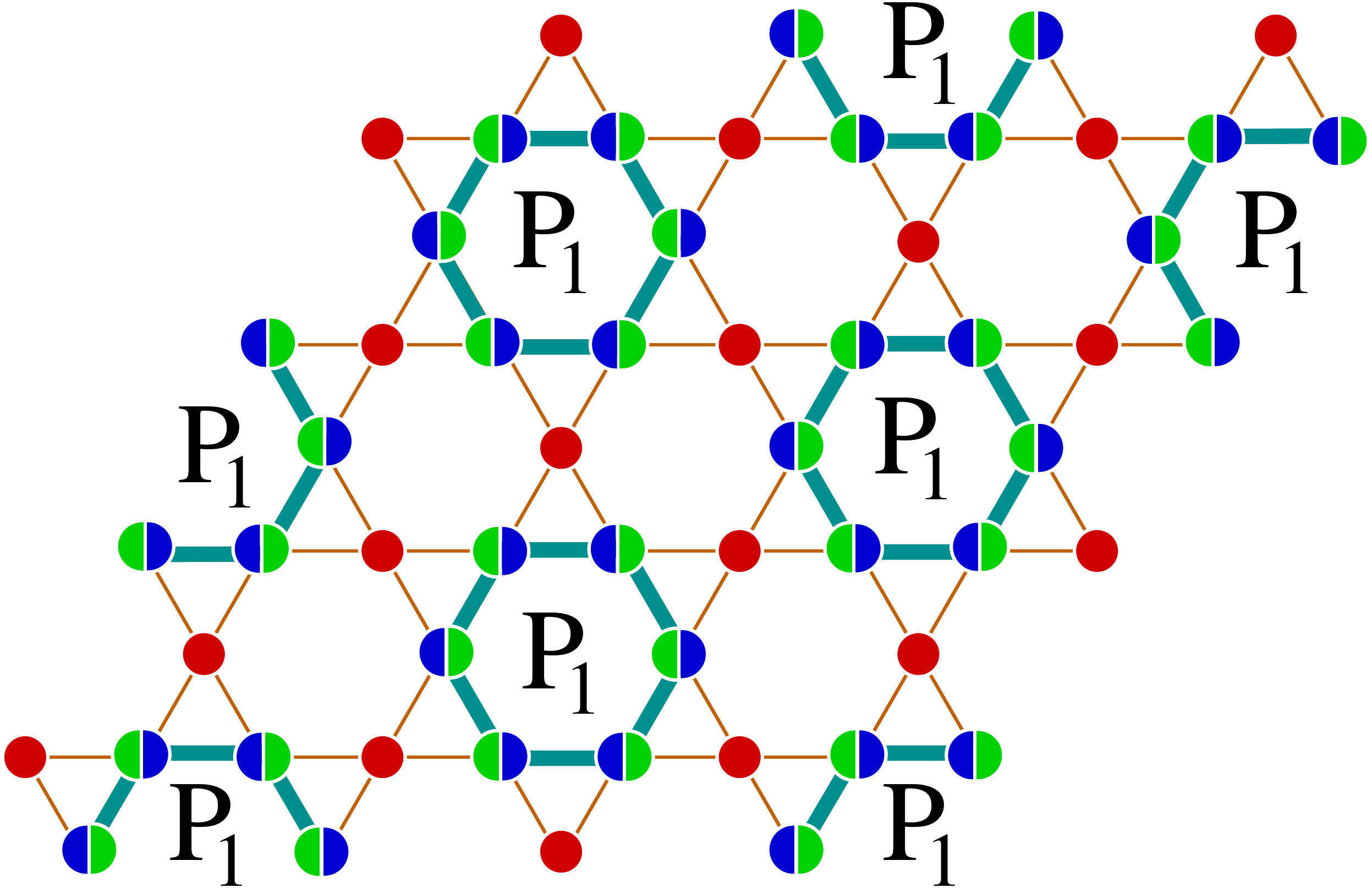}
\\
\vspace{5mm}
\includegraphics[width=0.55\linewidth]{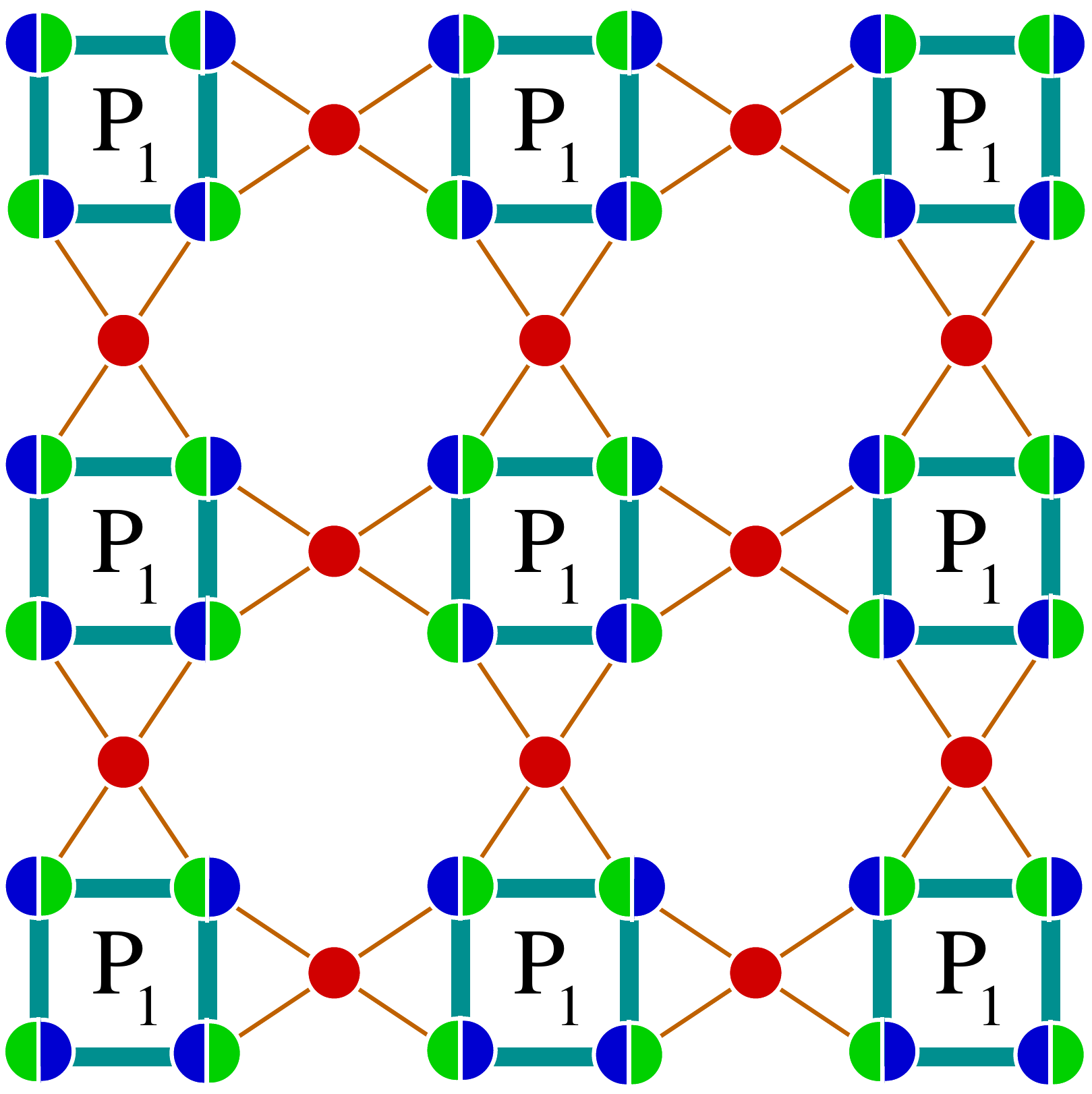}
\caption{(Color online) Many-body ground state wavefunction for magnons represented in a three-coloring basis. 
The top panel shows the case of 4 magnons (bosons) on the $4\times2\times3$ torus. 
The construction generalizes to $L$ magnons on the $L\times2\times3$ lattice i.e. 1/6 filling. Each magnon is confined to a strip 
and the many-body wavefunction is simply a product state of corresponding 
RCLs. Similar constructions apply at 1/9 filling to the infinite kagome and any finite cluster that accommodates the 
$\sqrt{3}\times\sqrt{3}$ pattern (middle panel). For 1/6 filling the construction also generalizes to two dimensions on the 
``squagome" lattice built up of triangular motifs (lowest panel).
\label{fig:color_map1} 
}
\end{figure}
	
Now that we have developed the connection between RCLs and localized and topological magnons, we will explicitly construct 
many-body solutions which minimize both $H_{XXZ0}$ and $H_{zz}$, for certain cases of net magnetization. 
We begin with the case of a narrow kagome torus with dimensions $L_x \times (L_y=2) \times 3$. 
For $L_x=4$, we show in Fig.~\ref{fig:color_map1} (top panel) 
that the RCLs are "stripes" (blue-green local motifs) on which the closely-packed localized magnons reside. 
Since each RCL is associated with a winding loop of 4 sites (along with 2 other padded sites) 
and each such motif contributes a single magnon or hard-core boson, the filling is exactly $1/6$. 
At this filling, denoted by $f$, the exact many body wavefunction is therefore a product state on these local motifs,
\begin{equation}
	|\psi \rangle = P_{Nf} \Big( \prod_{m = \text{motif}} | \text{RCL}_m \rangle \otimes \prod_{o = \text{other}} |r_o \rangle \Big)
\end{equation}
Since the magnons are never located on neighboring sites, due to the zero amplitude red sites 
(as indicated in the Fig.~\ref{fig:color_map1}) they completely avoid nearest neighbor density-density interactions 
(See Eq.~\eqref{eq:HXXZ_rewrite0}) thereby minimizing $H_{zz}$.
Thus, this wavefunction is the exact ground state for arbitrary repulsive interactions ($J_z \geq 0$). 
This closely-packed construction has been noted earlier in the literature in the $``+/-"$ magnon language~\cite{Richter_magnons, 
Zhitomirsky_Tsunetsugu, Richter_lectures, Schmidt_2002}. However, since the wavefunction is also a product of RCLs, 
the wavefunction has an exact representation in a basis of valid three-colorings, 
it also becomes the ground state for any $J_z \geq -1/2$, \textit{starting now}
from Eq.~\eqref{eq:XXZ0plusJz} via its hard-core boson counterpart.
The RCL is thus able to localize down spins (or particles) 
on motifs (eg. local hexagons or topological loops) and keeps them apart. 

This idea of constructing single magnon wavefunctions and the extension
to many body wavefunction generally applies to many other lattices, fillings and 
tilings (choices of motifs). For example, for $1/9$ filling, the idea generalizes to the 
infinite kagome and on any finite cluster that accommodates the $\sqrt{3}\times\sqrt{3}$ pattern. 
This includes the $36d$ cluster and certain quasi one-dimensional cylinders~\footnote{
The $\sqrt{3}\times\sqrt{3}$ pattern can fit on cylinders of width that are multiples of 4 but not multiples of 6 by shifting $\pm 2$ lattice constants 
in the periodic boundary condition along the cylinder circumference. These are $YC4+2$, $YC8-2$, $YC16-2$, $YC20+2$, etc. in the notation used in the DMRG literature. 
For cylinders of width that are multiples of 6, no such shifts are required}. 
Each magnon is now confined to a local hexagon and using the formalism of projected RCLs, the 
many body wavefunction is simply a product state of RCLs and color red ($a$) 
on sites that do not belong to the hexagonal RCLs. Since the tiling of RCLs can be done in three distinct ways (due to the 
three-fold symmetry of the $\sqrt{3}\times\sqrt{3}$ pattern), our construction yields a 
three-fold degenerate ground state solution. 

For $1/6$ filling, we may extend the above exact solutions to the two-dimensional "squagome" lattice, now 
using the motifs shown in Fig~\ref{fig:color_map1} (bottom panel). 
Each motif is once again associated with an RCL, and because of the $S_z$ or number projection operation, the 
intermediate sites between the magnons have zero amplitudes, with the magnon or boson residing on the square plaquettes.    

This analysis also immediately gives the ground state in the coloring basis for any lower density.  
If a wavefunction with $k$ RCLs is the ground state
of Eq.~\eqref{eq:XXZ0plusJz}, then the wavefunction obtained by replacing any subset of the $k$ RCLs by ICLs 
is still a ground state. In the thermodynamic limit, these fill in all
lower densities. In particular, this means that these phases extend to the quantum critical point at  $J_z=-1/2$ on the kagome for all filling $\leq 1/9$. 
 
Thus, we have shown that several low magnon (particle) density/high magnetization solutions can be exactly constructed from three-coloring states. 
In each individual example presented, the wavefunction constructed 
minimizes both $H_{XXZ0}$ and $H_{zz}$, and hence is the \textit{exact} ground state wavefunction for any $J_z \geq -1/2$. 
Said differently, the magnons confined to their individual motifs (strip, hexagon etc.) completely avoid repulsion ($J_z >0$) 
at low density and minimize their kinetic energy by staying localized. However the color-magnon transformation shows that even 
under attractive interactions, the localized magnons do not immediately condense - rather there is a \textit{critical attraction} 
strength ($J_z=-1/2$) which is needed for this to happen. While this result is true and mathematically rigorous only at low density, where the magnons form a crystal, 
a natural question that arises is whether the coloring manifold is responsible for the origin of the spin liquid ground state, expected at one-sixth 
(2/3 magnetization)~\cite{Kumar2014,KumarChanglani2016} and half filling (zero magnetization). 

\section{DMRG for $H_{XXZ}$ for the zero magnetization sector}
\label{sec:dmrg}
In lieu of such an understanding, we now turn our attention to a numerical study in the case of half filling ($S_z=0$) 
where the ground state does not have an exact three-coloring representation. A previous DMRG study~\cite{He_PRL} argued that the spin liquid at $J_z=1$ (the Heisenberg point) 
adiabatically continues both to the $J_z=0$ and $J_z \gg 1$ limits and an ED study~\cite{Lauchli_48} 
showed a remarkable similarity in the low energy spectrum from $J_z=0$ to $J_z=1$. In addition, 
another ED study on $36$ sites strongly suggested adiabatic continuity for all $J_z \gtrsim -0.4$ (and possibly beyond)~\cite{Changlani_PRL}. 
Here, we extend these results by performing large-scale DMRG calculations (using ITensor~\cite{Itensor}) 
for the $J_z<0$ case; these results support the finding that the spin liquid phase extends to the $J_z=-1/2$ point~\cite{Changlani_PRL}~\footnote{
This argument does not account for the possibility that the $XXZ$ model for any $J_z\geq-1/2$ is potentially 
part of a critical line~\cite{Changlani_PRL}}.
\begin{figure}
\centering
\includegraphics[width=0.95\linewidth]{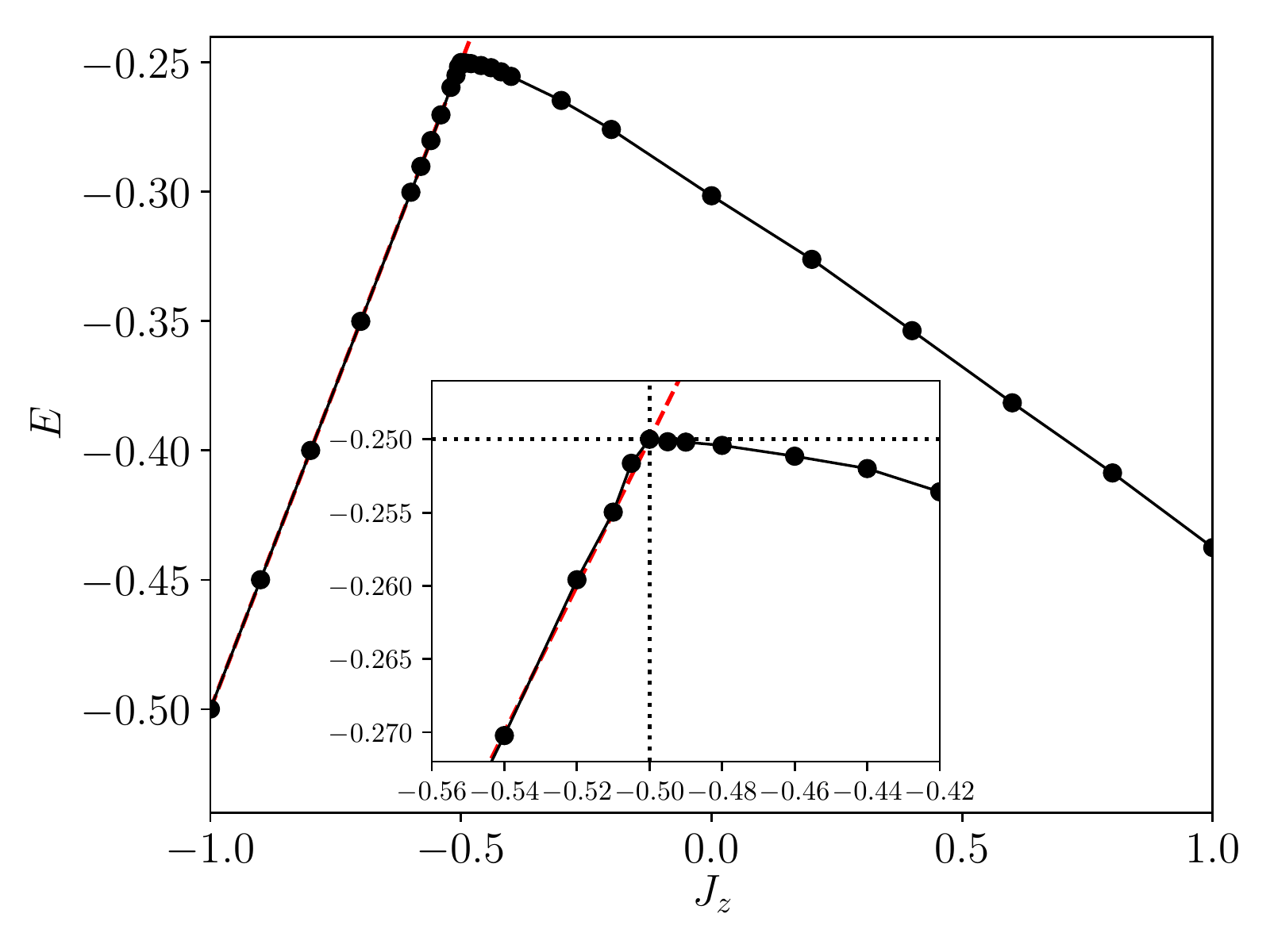}
\caption{(Color online) Ground state energy per site from DMRG for the XC-8 
cylinder in the limit of infinite length for the 
range $-1 \leq J_z \leq 1$. The red dashed line indicates the energy ($=J_z/2$) of pure ferromagnetic states. 
The inset zooms into a narrow range around $J_z = -1/2$. The errorbars are presented but smaller than the symbol sizes. The dotted 
lines in the inset indicate the exact energy $-1/4$ at $J_z=-1/2$.}
\label{fig:energies_XC8} 
\end{figure}

\begin{figure}
\centering
\includegraphics[width=0.95\linewidth]{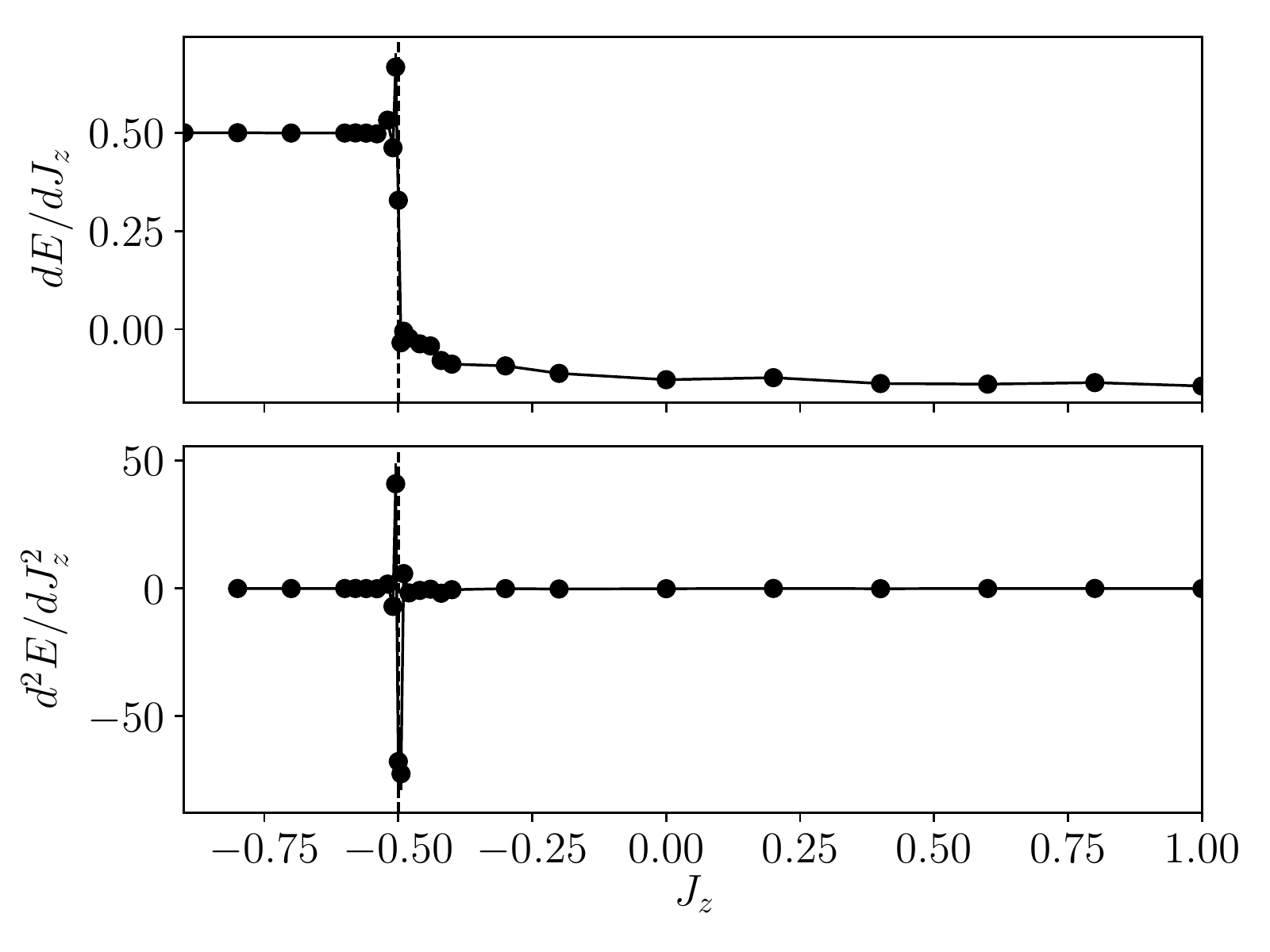}
\caption{First and second derivative of the energy per site as a function of $J_z$. 
The errorbars are presented but smaller than the symbol sizes. The discontinuity in the first derivative and the peak 
in the second derivative at $J_z=-1/2$ signal the occurrence of a quantum phase transition.}
\label{fig:energy_derivatives} 
\end{figure}

\begin{figure}
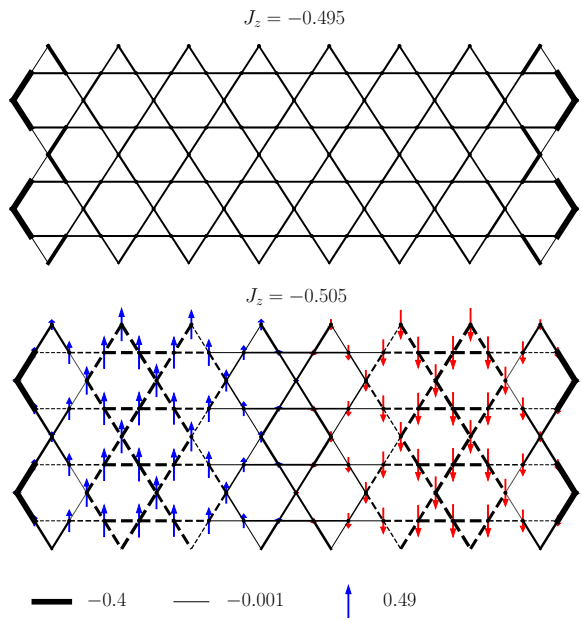

\centering
\includegraphics[width=0.95\linewidth]{{{./Jz-0.495_8x2.out.sz}}}
\includegraphics[width=0.95\linewidth]{{{./Jz-0.505_8x2.out.sz}}}
\caption{(Color online) Spatial profile of spin moments $\langle S^{z}_i\rangle$ and 
valence bond energies $\langle\mathbf{S}_i\cdot\mathbf{S}_j\rangle$ on a representative 
XC-8 cylinder for $J_z=-0.495$ and $J_z=-0.505$. The maximum spin moment for $J_z=-0.495$ is $\sim 5\times10^{-4}$.
The solid~(dashed) bonds represent the negative~(positive) valence bond energies.}
\label{fig:spin_bond} 
\end{figure}

We study the zero-magnetization ground states in a wide range of $J_z$, from $J_z=5$ to $J_z=-1$. 
To better focus on the $J_z=-1/2$ point, we have shown the results only up to $J_z=1$; 
the ground state changes smoothly with no signs of a phase transition between $J_z=1$ and $J_z=5$.
We focus on the $XC8$ cylindrical geometry (which is depicted in Fig.~\ref{fig:spin_bond}) and keep the number of states up to $7000$. 
The total energy has been extrapolated to infinite-length; our extrapolated results 
are shown in Fig.~\ref{fig:energies_XC8} as a function of $J_z$. 

For the region of $J_z<-1/2$ the ground state is ferromagnetic (albeit phase-separated due to the $S_z=0$ constraint), 
and thus the ground state energy for this region equals $J_z/2$, as 
indicated by the good agreement between our DMRG data points and the red dashed line.
On going from $J_z=1$ to $J_z=0$ the energy increases monotonically and smoothly, indicating an 
absence of a phase transition in this region, consistent with Ref.~\cite{He_PRL}. 
Importantly, this smooth monotonic behavior continues across $J_z=0$ and a kink is seen only at (or close to) $J_z=-1/2$, 
strongly suggesting that the exactly solvable point is a transition point between spin liquid and ferromagnetic states. 
Evidence for such a transition is further clarified by monitoring the first and second derivatives of energy, 
shown in Fig.~\ref{fig:energy_derivatives}; the first derivative has a discontinuity and the second derivative has a peak at $J_z=-1/2$.

In practice, the DMRG simulations were found to get stuck in valence bond solid states
or metastable states with edge spins. We thus had to run different random 
initial states to converge to the lowest energy spin liquid state.
We found that the convergence is particularly difficult around $J_z=-0.4$, 
which may suggest the need for further detailed future studies in this 
region with larger systems and different geometries. 
For $J_z<-1/2$ we started our DMRG calculations with the two domain state.
Not doing so led to more ferromagnetic domains with slightly higher energy than
the two domain solution. Magnetic pinning field is also applied to further stabilize the states in the region 
$-0.52 \leq J_z< -0.5$
close to the transition point.

Fig.~\ref{fig:spin_bond} shows local spin and valence bond order parameters at 
two representative points $J_z=-0.495$ and $J_z=-0.505$, close to the transition point. 
Clearly, for $J_z=-0.495$ there is no local order (for the order parameters measured) 
and for $J_z=-0.505$ a ferromagnetic state is stabilized; domains are observed as the system prefers to phase-separate to maintain the $S_z=0$ constraint.

\section{Conclusion}
\label{sec:discuss}
In summary, we have explored properties of quantum three-coloring states and developed an exact one-to-one correspondence between 
quantum three-colors and the localized and topological magnons that make up the flat band modes on the kagome lattice. While both perspectives and concepts have existed in the 
literature in various forms (classical three-colorings, quantum magnons), our work makes their connection concrete for the quantum case 
and generalizes it to both the single and multi magnon case. It is no coincidence that the two color loops in a 
three-coloring state, and the magnon modes in the kagome flat band have geometrical similarities; our work shows why this is the case. 

Extending this connection, we have expressed exact many-body ground state wavefunctions for special high magnetizations (or low fillings in the bosonic language) 
in a three coloring basis; this proves their validity for all $J_z\geq-1/2$ showing the equivalence of the $XY$ and Ising regimes, 
for these magnetization sectors. Using the color-magnon transformation, our results extend the range of validity of 
exact solutions which have been argued to hold for $J_z \geq 0$ (repulsive case in the boson language), 
to $-1/2\leq J_z \leq0$ (\emph{attractive} case). We have also highlighted the important role and subtleties of number projection at low fillings. 
For the case of half filling/zero magnetization, our numerical DMRG calculations suggest that the physics of the Heisenberg point 
is crucially connected to the $J_z=-1/2$ point. 

Finally, we note that in the present work, we have only considered the cases where a macroscopic superposition of 
three colorings describes product states in the magnon basis (these are incidentally also a subset of correlator product states~\cite{Changlani_CPS}). 
However, it is natural to ask whether and/or how can one map
a highly entangled state from the Ising or magnon basis to the three-color basis. In addition, 
the three-colorings present an attractive possibility of explaining the 
large number (exponentially scaling with system size) of singlets seen in the low energy spectrum in exact diagonalizations~\cite{Waldtmann_Lhuillier, Mila_kagome}.
We hope to address these and related questions in the future.

\section{Acknowledgements} 
HJC and BKC thank E. Fradkin, D. Kochkov and K. Kumar for an earlier collaboration. We thank O. Vafek, S. Sachdev, O.Tchernyshyov, 
P. Nikolic, V. Dobrosavljevic, F. Verstraete, A. Ralko, Y-C.He and M. Lawler for their encouragement and for useful discussions. 
This work was supported through the Institute for Quantum Matter at Johns Hopkins University, by the 
U.S. Department of Energy, Division of Basic Energy 
Sciences, Grant DE-FG02-08ER46544. HJC acknowledges start up funds at Florida State University. 
SP acknowledges the support (17IRCCSG011) of IRCC, IIT Bombay. 
We gratefully acknowledge the Johns Hopkins Homewood High Performance Cluster (HHPC) 
and the Maryland Advanced Research Computing Center (MARCC), funded by the State of Maryland, for computing resources. 
This research is also part of the BlueWaters sustained petascale computing project, which is supported by the National Science Foundation 
(award numbers OCI-0725070 and ACI-1238993) and the State of Illinois.

\bibliographystyle{apsrev4-1}
\bibliography{refs}

\begin{thebibliography}{46}%
\makeatletter
\providecommand \@ifxundefined [1]{%
 \@ifx{#1\undefined}
}%
\providecommand \@ifnum [1]{%
 \ifnum #1\expandafter \@firstoftwo
 \else \expandafter \@secondoftwo
 \fi
}%
\providecommand \@ifx [1]{%
 \ifx #1\expandafter \@firstoftwo
 \else \expandafter \@secondoftwo
 \fi
}%
\providecommand \natexlab [1]{#1}%
\providecommand \enquote  [1]{``#1''}%
\providecommand \bibnamefont  [1]{#1}%
\providecommand \bibfnamefont [1]{#1}%
\providecommand \citenamefont [1]{#1}%
\providecommand \href@noop [0]{\@secondoftwo}%
\providecommand \href [0]{\begingroup \@sanitize@url \@href}%
\providecommand \@href[1]{\@@startlink{#1}\@@href}%
\providecommand \@@href[1]{\endgroup#1\@@endlink}%
\providecommand \@sanitize@url [0]{\catcode `\\12\catcode `\$12\catcode
  `\&12\catcode `\#12\catcode `\^12\catcode `\_12\catcode `\%12\relax}%
\providecommand \@@startlink[1]{}%
\providecommand \@@endlink[0]{}%
\providecommand \url  [0]{\begingroup\@sanitize@url \@url }%
\providecommand \@url [1]{\endgroup\@href {#1}{\urlprefix }}%
\providecommand \urlprefix  [0]{URL }%
\providecommand \Eprint [0]{\href }%
\providecommand \doibase [0]{http://dx.doi.org/}%
\providecommand \selectlanguage [0]{\@gobble}%
\providecommand \bibinfo  [0]{\@secondoftwo}%
\providecommand \bibfield  [0]{\@secondoftwo}%
\providecommand \translation [1]{[#1]}%
\providecommand \BibitemOpen [0]{}%
\providecommand \bibitemStop [0]{}%
\providecommand \bibitemNoStop [0]{.\EOS\space}%
\providecommand \EOS [0]{\spacefactor3000\relax}%
\providecommand \BibitemShut  [1]{\csname bibitem#1\endcsname}%
\let\auto@bib@innerbib\@empty
\bibitem [{\citenamefont {Shores}\ \emph {et~al.}(2005)\citenamefont {Shores},
  \citenamefont {Nytko}, \citenamefont {Bartlett},\ and\ \citenamefont
  {Nocera}}]{Nocera_Kagome_2005}%
  \BibitemOpen
  \bibfield  {author} {\bibinfo {author} {\bibfnamefont {M.~P.}\ \bibnamefont
  {Shores}}, \bibinfo {author} {\bibfnamefont {E.~A.}\ \bibnamefont {Nytko}},
  \bibinfo {author} {\bibfnamefont {B.~M.}\ \bibnamefont {Bartlett}}, \ and\
  \bibinfo {author} {\bibfnamefont {D.~G.}\ \bibnamefont {Nocera}},\ }\href
  {\doibase 10.1021/ja053891p} {\bibfield  {journal} {\bibinfo  {journal}
  {Journal of the American Chemical Society}\ }\textbf {\bibinfo {volume}
  {127}},\ \bibinfo {pages} {13462} (\bibinfo {year} {2005})},\ \bibinfo {note}
  {pMID: 16190686},\ \Eprint
  {http://arxiv.org/abs/https://doi.org/10.1021/ja053891p}
  {https://doi.org/10.1021/ja053891p} \BibitemShut {NoStop}%
\bibitem [{\citenamefont {Helton}\ \emph {et~al.}(2007)\citenamefont {Helton},
  \citenamefont {Matan}, \citenamefont {Shores}, \citenamefont {Nytko},
  \citenamefont {Bartlett}, \citenamefont {Yoshida}, \citenamefont {Takano},
  \citenamefont {Suslov}, \citenamefont {Qiu}, \citenamefont {Chung},
  \citenamefont {Nocera},\ and\ \citenamefont {Lee}}]{Helton_2007}%
  \BibitemOpen
  \bibfield  {author} {\bibinfo {author} {\bibfnamefont {J.~S.}\ \bibnamefont
  {Helton}}, \bibinfo {author} {\bibfnamefont {K.}~\bibnamefont {Matan}},
  \bibinfo {author} {\bibfnamefont {M.~P.}\ \bibnamefont {Shores}}, \bibinfo
  {author} {\bibfnamefont {E.~A.}\ \bibnamefont {Nytko}}, \bibinfo {author}
  {\bibfnamefont {B.~M.}\ \bibnamefont {Bartlett}}, \bibinfo {author}
  {\bibfnamefont {Y.}~\bibnamefont {Yoshida}}, \bibinfo {author} {\bibfnamefont
  {Y.}~\bibnamefont {Takano}}, \bibinfo {author} {\bibfnamefont
  {A.}~\bibnamefont {Suslov}}, \bibinfo {author} {\bibfnamefont
  {Y.}~\bibnamefont {Qiu}}, \bibinfo {author} {\bibfnamefont {J.-H.}\
  \bibnamefont {Chung}}, \bibinfo {author} {\bibfnamefont {D.~G.}\ \bibnamefont
  {Nocera}}, \ and\ \bibinfo {author} {\bibfnamefont {Y.~S.}\ \bibnamefont
  {Lee}},\ }\href {\doibase 10.1103/PhysRevLett.98.107204} {\bibfield
  {journal} {\bibinfo  {journal} {Phys. Rev. Lett.}\ }\textbf {\bibinfo
  {volume} {98}},\ \bibinfo {pages} {107204} (\bibinfo {year}
  {2007})}\BibitemShut {NoStop}%
\bibitem [{\citenamefont {Mendels}\ and\ \citenamefont
  {Bert}(2016)}]{Mendels_Bert}%
  \BibitemOpen
  \bibfield  {author} {\bibinfo {author} {\bibfnamefont {P.}~\bibnamefont
  {Mendels}}\ and\ \bibinfo {author} {\bibfnamefont {F.}~\bibnamefont {Bert}},\
  }\href {\doibase https://doi.org/10.1016/j.crhy.2015.12.001} {\bibfield
  {journal} {\bibinfo  {journal} {Comptes Rendus Physique}\ }\textbf {\bibinfo
  {volume} {17}},\ \bibinfo {pages} {455 } (\bibinfo {year}
  {2016})}\BibitemShut {NoStop}%
\bibitem [{\citenamefont {Jeschke}\ \emph {et~al.}(2013)\citenamefont
  {Jeschke}, \citenamefont {Salvat-Pujol},\ and\ \citenamefont
  {Valenti}}]{Jeschke_2013}%
  \BibitemOpen
  \bibfield  {author} {\bibinfo {author} {\bibfnamefont {H.~O.}\ \bibnamefont
  {Jeschke}}, \bibinfo {author} {\bibfnamefont {F.}~\bibnamefont
  {Salvat-Pujol}}, \ and\ \bibinfo {author} {\bibfnamefont {R.}~\bibnamefont
  {Valenti}},\ }\href {\doibase 10.1103/PhysRevB.88.075106} {\bibfield
  {journal} {\bibinfo  {journal} {Phys. Rev. B}\ }\textbf {\bibinfo {volume}
  {88}},\ \bibinfo {pages} {075106} (\bibinfo {year} {2013})}\BibitemShut
  {NoStop}%
\bibitem [{\citenamefont {Zeng}\ and\ \citenamefont {Elser}(1990)}]{ZengElser}%
  \BibitemOpen
  \bibfield  {author} {\bibinfo {author} {\bibfnamefont {C.}~\bibnamefont
  {Zeng}}\ and\ \bibinfo {author} {\bibfnamefont {V.}~\bibnamefont {Elser}},\
  }\href {\doibase 10.1103/PhysRevB.42.8436} {\bibfield  {journal} {\bibinfo
  {journal} {Phys. Rev. B}\ }\textbf {\bibinfo {volume} {42}},\ \bibinfo
  {pages} {8436} (\bibinfo {year} {1990})}\BibitemShut {NoStop}%
\bibitem [{\citenamefont {Ran}\ \emph {et~al.}(2007)\citenamefont {Ran},
  \citenamefont {Hermele}, \citenamefont {Lee},\ and\ \citenamefont
  {Wen}}]{Wen_Kagome}%
  \BibitemOpen
  \bibfield  {author} {\bibinfo {author} {\bibfnamefont {Y.}~\bibnamefont
  {Ran}}, \bibinfo {author} {\bibfnamefont {M.}~\bibnamefont {Hermele}},
  \bibinfo {author} {\bibfnamefont {P.~A.}\ \bibnamefont {Lee}}, \ and\
  \bibinfo {author} {\bibfnamefont {X.-G.}\ \bibnamefont {Wen}},\ }\href
  {\doibase 10.1103/PhysRevLett.98.117205} {\bibfield  {journal} {\bibinfo
  {journal} {Phys. Rev. Lett.}\ }\textbf {\bibinfo {volume} {98}},\ \bibinfo
  {pages} {117205} (\bibinfo {year} {2007})}\BibitemShut {NoStop}%
\bibitem [{\citenamefont {Yan}\ \emph {et~al.}(2011)\citenamefont {Yan},
  \citenamefont {Huse},\ and\ \citenamefont {White}}]{White_Kagome}%
  \BibitemOpen
  \bibfield  {author} {\bibinfo {author} {\bibfnamefont {S.}~\bibnamefont
  {Yan}}, \bibinfo {author} {\bibfnamefont {D.~A.}\ \bibnamefont {Huse}}, \
  and\ \bibinfo {author} {\bibfnamefont {S.~R.}\ \bibnamefont {White}},\ }\href
  {\doibase 10.1126/science.1201080} {\bibfield  {journal} {\bibinfo  {journal}
  {Science}\ }\textbf {\bibinfo {volume} {332}},\ \bibinfo {pages} {1173}
  (\bibinfo {year} {2011})}\BibitemShut {NoStop}%
\bibitem [{\citenamefont {Depenbrock}\ \emph {et~al.}(2012)\citenamefont
  {Depenbrock}, \citenamefont {McCulloch},\ and\ \citenamefont
  {Schollw\"ock}}]{Depenbrock_Kagome}%
  \BibitemOpen
  \bibfield  {author} {\bibinfo {author} {\bibfnamefont {S.}~\bibnamefont
  {Depenbrock}}, \bibinfo {author} {\bibfnamefont {I.~P.}\ \bibnamefont
  {McCulloch}}, \ and\ \bibinfo {author} {\bibfnamefont {U.}~\bibnamefont
  {Schollw\"ock}},\ }\href {\doibase 10.1103/PhysRevLett.109.067201} {\bibfield
   {journal} {\bibinfo  {journal} {Phys. Rev. Lett.}\ }\textbf {\bibinfo
  {volume} {109}},\ \bibinfo {pages} {067201} (\bibinfo {year}
  {2012})}\BibitemShut {NoStop}%
\bibitem [{\citenamefont {Iqbal}\ \emph {et~al.}(2013)\citenamefont {Iqbal},
  \citenamefont {Becca}, \citenamefont {Sorella},\ and\ \citenamefont
  {Poilblanc}}]{Iqbal_Kagome}%
  \BibitemOpen
  \bibfield  {author} {\bibinfo {author} {\bibfnamefont {Y.}~\bibnamefont
  {Iqbal}}, \bibinfo {author} {\bibfnamefont {F.}~\bibnamefont {Becca}},
  \bibinfo {author} {\bibfnamefont {S.}~\bibnamefont {Sorella}}, \ and\
  \bibinfo {author} {\bibfnamefont {D.}~\bibnamefont {Poilblanc}},\ }\href
  {\doibase 10.1103/PhysRevB.87.060405} {\bibfield  {journal} {\bibinfo
  {journal} {Phys. Rev. B}\ }\textbf {\bibinfo {volume} {87}},\ \bibinfo
  {pages} {060405} (\bibinfo {year} {2013})}\BibitemShut {NoStop}%
\bibitem [{\citenamefont {Jiang}\ \emph {et~al.}(2012)\citenamefont {Jiang},
  \citenamefont {Wang},\ and\ \citenamefont {Balents}}]{Jiang_Balents}%
  \BibitemOpen
  \bibfield  {author} {\bibinfo {author} {\bibfnamefont {H.-C.}\ \bibnamefont
  {Jiang}}, \bibinfo {author} {\bibfnamefont {Z.}~\bibnamefont {Wang}}, \ and\
  \bibinfo {author} {\bibfnamefont {L.}~\bibnamefont {Balents}},\ }\href
  {\doibase 10.1038/nphys2465} {\bibfield  {journal} {\bibinfo  {journal} {Nat.
  Phys.}\ }\textbf {\bibinfo {volume} {8}},\ \bibinfo {pages} {902} (\bibinfo
  {year} {2012})}\BibitemShut {NoStop}%
\bibitem [{\citenamefont {Tay}\ and\ \citenamefont
  {Motrunich}(2011)}]{Tay_Motrunich}%
  \BibitemOpen
  \bibfield  {author} {\bibinfo {author} {\bibfnamefont {T.}~\bibnamefont
  {Tay}}\ and\ \bibinfo {author} {\bibfnamefont {O.~I.}\ \bibnamefont
  {Motrunich}},\ }\href {\doibase 10.1103/PhysRevB.84.020404} {\bibfield
  {journal} {\bibinfo  {journal} {Phys. Rev. B}\ }\textbf {\bibinfo {volume}
  {84}},\ \bibinfo {pages} {020404} (\bibinfo {year} {2011})}\BibitemShut
  {NoStop}%
\bibitem [{\citenamefont {He}\ \emph {et~al.}(2017)\citenamefont {He},
  \citenamefont {Zaletel}, \citenamefont {Oshikawa},\ and\ \citenamefont
  {Pollmann}}]{He_Zaletel_Kagome}%
  \BibitemOpen
  \bibfield  {author} {\bibinfo {author} {\bibfnamefont {Y.-C.}\ \bibnamefont
  {He}}, \bibinfo {author} {\bibfnamefont {M.~P.}\ \bibnamefont {Zaletel}},
  \bibinfo {author} {\bibfnamefont {M.}~\bibnamefont {Oshikawa}}, \ and\
  \bibinfo {author} {\bibfnamefont {F.}~\bibnamefont {Pollmann}},\ }\href
  {\doibase 10.1103/PhysRevX.7.031020} {\bibfield  {journal} {\bibinfo
  {journal} {Phys. Rev. X}\ }\textbf {\bibinfo {volume} {7}},\ \bibinfo {pages}
  {031020} (\bibinfo {year} {2017})}\BibitemShut {NoStop}%
\bibitem [{\citenamefont {Liao}\ \emph {et~al.}(2017)\citenamefont {Liao},
  \citenamefont {Xie}, \citenamefont {Chen}, \citenamefont {Liu}, \citenamefont
  {Xie}, \citenamefont {Huang}, \citenamefont {Normand},\ and\ \citenamefont
  {Xiang}}]{Normand_Xiang}%
  \BibitemOpen
  \bibfield  {author} {\bibinfo {author} {\bibfnamefont {H.~J.}\ \bibnamefont
  {Liao}}, \bibinfo {author} {\bibfnamefont {Z.~Y.}\ \bibnamefont {Xie}},
  \bibinfo {author} {\bibfnamefont {J.}~\bibnamefont {Chen}}, \bibinfo {author}
  {\bibfnamefont {Z.~Y.}\ \bibnamefont {Liu}}, \bibinfo {author} {\bibfnamefont
  {H.~D.}\ \bibnamefont {Xie}}, \bibinfo {author} {\bibfnamefont {R.~Z.}\
  \bibnamefont {Huang}}, \bibinfo {author} {\bibfnamefont {B.}~\bibnamefont
  {Normand}}, \ and\ \bibinfo {author} {\bibfnamefont {T.}~\bibnamefont
  {Xiang}},\ }\href {\doibase 10.1103/PhysRevLett.118.137202} {\bibfield
  {journal} {\bibinfo  {journal} {Phys. Rev. Lett.}\ }\textbf {\bibinfo
  {volume} {118}},\ \bibinfo {pages} {137202} (\bibinfo {year}
  {2017})}\BibitemShut {NoStop}%
\bibitem [{\citenamefont {Messio}\ \emph {et~al.}(2012)\citenamefont {Messio},
  \citenamefont {Bernu},\ and\ \citenamefont {Lhuillier}}]{Messio}%
  \BibitemOpen
  \bibfield  {author} {\bibinfo {author} {\bibfnamefont {L.}~\bibnamefont
  {Messio}}, \bibinfo {author} {\bibfnamefont {B.}~\bibnamefont {Bernu}}, \
  and\ \bibinfo {author} {\bibfnamefont {C.}~\bibnamefont {Lhuillier}},\ }\href
  {\doibase 10.1103/PhysRevLett.108.207204} {\bibfield  {journal} {\bibinfo
  {journal} {Phys. Rev. Lett.}\ }\textbf {\bibinfo {volume} {108}},\ \bibinfo
  {pages} {207204} (\bibinfo {year} {2012})}\BibitemShut {NoStop}%
\bibitem [{\citenamefont {Hao}\ and\ \citenamefont
  {Tchernyshyov}(2009)}]{Hao_Tchernyshyov}%
  \BibitemOpen
  \bibfield  {author} {\bibinfo {author} {\bibfnamefont {Z.}~\bibnamefont
  {Hao}}\ and\ \bibinfo {author} {\bibfnamefont {O.}~\bibnamefont
  {Tchernyshyov}},\ }\href {\doibase 10.1103/PhysRevLett.103.187203} {\bibfield
   {journal} {\bibinfo  {journal} {Phys. Rev. Lett.}\ }\textbf {\bibinfo
  {volume} {103}},\ \bibinfo {pages} {187203} (\bibinfo {year}
  {2009})}\BibitemShut {NoStop}%
\bibitem [{\citenamefont {Ralko}\ \emph {et~al.}(2018)\citenamefont {Ralko},
  \citenamefont {Mila},\ and\ \citenamefont {Rousochatzakis}}]{Ralko_Mila}%
  \BibitemOpen
  \bibfield  {author} {\bibinfo {author} {\bibfnamefont {A.}~\bibnamefont
  {Ralko}}, \bibinfo {author} {\bibfnamefont {F.}~\bibnamefont {Mila}}, \ and\
  \bibinfo {author} {\bibfnamefont {I.}~\bibnamefont {Rousochatzakis}},\ }\href
  {\doibase 10.1103/PhysRevB.97.104401} {\bibfield  {journal} {\bibinfo
  {journal} {Phys. Rev. B}\ }\textbf {\bibinfo {volume} {97}},\ \bibinfo
  {pages} {104401} (\bibinfo {year} {2018})}\BibitemShut {NoStop}%
\bibitem [{\citenamefont {Changlani}\ \emph {et~al.}(2018)\citenamefont
  {Changlani}, \citenamefont {Kochkov}, \citenamefont {Kumar}, \citenamefont
  {Clark},\ and\ \citenamefont {Fradkin}}]{Changlani_PRL}%
  \BibitemOpen
  \bibfield  {author} {\bibinfo {author} {\bibfnamefont {H.~J.}\ \bibnamefont
  {Changlani}}, \bibinfo {author} {\bibfnamefont {D.}~\bibnamefont {Kochkov}},
  \bibinfo {author} {\bibfnamefont {K.}~\bibnamefont {Kumar}}, \bibinfo
  {author} {\bibfnamefont {B.~K.}\ \bibnamefont {Clark}}, \ and\ \bibinfo
  {author} {\bibfnamefont {E.}~\bibnamefont {Fradkin}},\ }\href {\doibase
  10.1103/PhysRevLett.120.117202} {\bibfield  {journal} {\bibinfo  {journal}
  {Phys. Rev. Lett.}\ }\textbf {\bibinfo {volume} {120}},\ \bibinfo {pages}
  {117202} (\bibinfo {year} {2018})}\BibitemShut {NoStop}%
\bibitem [{\citenamefont {{Li}}(2011)}]{Tao_Li_critical1}%
  \BibitemOpen
  \bibfield  {author} {\bibinfo {author} {\bibfnamefont {T.}~\bibnamefont
  {{Li}}},\ }\href@noop {} {\bibfield  {journal} {\bibinfo  {journal} {ArXiv
  e-prints}\ } (\bibinfo {year} {2011})},\ \Eprint
  {http://arxiv.org/abs/1106.6134} {arXiv:1106.6134 [cond-mat.str-el]}
  \BibitemShut {NoStop}%
\bibitem [{\citenamefont {Essafi}\ \emph {et~al.}(2016)\citenamefont {Essafi},
  \citenamefont {Benton},\ and\ \citenamefont {Jaubert}}]{Jaubert2016}%
  \BibitemOpen
  \bibfield  {author} {\bibinfo {author} {\bibfnamefont {K.}~\bibnamefont
  {Essafi}}, \bibinfo {author} {\bibfnamefont {O.}~\bibnamefont {Benton}}, \
  and\ \bibinfo {author} {\bibfnamefont {L.~D.~C.}\ \bibnamefont {Jaubert}},\
  }\href {http://dx.doi.org/10.1038/ncomms10297} {\bibfield  {journal}
  {\bibinfo  {journal} {Nat. Commun.}\ }\textbf {\bibinfo {volume} {7}},\
  \bibinfo {pages} {10297} (\bibinfo {year} {2016})}\BibitemShut {NoStop}%
\bibitem [{\citenamefont {Harris}\ \emph {et~al.}(1992)\citenamefont {Harris},
  \citenamefont {Kallin},\ and\ \citenamefont {Berlinsky}}]{Harris3color}%
  \BibitemOpen
  \bibfield  {author} {\bibinfo {author} {\bibfnamefont {A.~B.}\ \bibnamefont
  {Harris}}, \bibinfo {author} {\bibfnamefont {C.}~\bibnamefont {Kallin}}, \
  and\ \bibinfo {author} {\bibfnamefont {A.~J.}\ \bibnamefont {Berlinsky}},\
  }\href {\doibase 10.1103/PhysRevB.45.2899} {\bibfield  {journal} {\bibinfo
  {journal} {Phys. Rev. B}\ }\textbf {\bibinfo {volume} {45}},\ \bibinfo
  {pages} {2899} (\bibinfo {year} {1992})}\BibitemShut {NoStop}%
\bibitem [{\citenamefont {Henley}(2009)}]{Henley3color}%
  \BibitemOpen
  \bibfield  {author} {\bibinfo {author} {\bibfnamefont {C.~L.}\ \bibnamefont
  {Henley}},\ }\href {\doibase 10.1103/PhysRevB.80.180401} {\bibfield
  {journal} {\bibinfo  {journal} {Phys. Rev. B}\ }\textbf {\bibinfo {volume}
  {80}},\ \bibinfo {pages} {180401} (\bibinfo {year} {2009})}\BibitemShut
  {NoStop}%
\bibitem [{\citenamefont {Huse}\ and\ \citenamefont
  {Rutenberg}(1992)}]{Huse_Rutenberg}%
  \BibitemOpen
  \bibfield  {author} {\bibinfo {author} {\bibfnamefont {D.~A.}\ \bibnamefont
  {Huse}}\ and\ \bibinfo {author} {\bibfnamefont {A.~D.}\ \bibnamefont
  {Rutenberg}},\ }\href {\doibase 10.1103/PhysRevB.45.7536} {\bibfield
  {journal} {\bibinfo  {journal} {Phys. Rev. B}\ }\textbf {\bibinfo {volume}
  {45}},\ \bibinfo {pages} {7536} (\bibinfo {year} {1992})}\BibitemShut
  {NoStop}%
\bibitem [{\citenamefont {Chalker}\ \emph {et~al.}(1992)\citenamefont
  {Chalker}, \citenamefont {Holdsworth},\ and\ \citenamefont
  {Shender}}]{Chalker3color}%
  \BibitemOpen
  \bibfield  {author} {\bibinfo {author} {\bibfnamefont {J.~T.}\ \bibnamefont
  {Chalker}}, \bibinfo {author} {\bibfnamefont {P.~C.~W.}\ \bibnamefont
  {Holdsworth}}, \ and\ \bibinfo {author} {\bibfnamefont {E.~F.}\ \bibnamefont
  {Shender}},\ }\href {\doibase 10.1103/PhysRevLett.68.855} {\bibfield
  {journal} {\bibinfo  {journal} {Phys. Rev. Lett.}\ }\textbf {\bibinfo
  {volume} {68}},\ \bibinfo {pages} {855} (\bibinfo {year} {1992})}\BibitemShut
  {NoStop}%
\bibitem [{\citenamefont {Sachdev}(1992)}]{Sachdev92}%
  \BibitemOpen
  \bibfield  {author} {\bibinfo {author} {\bibfnamefont {S.}~\bibnamefont
  {Sachdev}},\ }\href {\doibase 10.1103/PhysRevB.45.12377} {\bibfield
  {journal} {\bibinfo  {journal} {Phys. Rev. B}\ }\textbf {\bibinfo {volume}
  {45}},\ \bibinfo {pages} {12377} (\bibinfo {year} {1992})}\BibitemShut
  {NoStop}%
\bibitem [{\citenamefont {von Delft}\ and\ \citenamefont
  {Henley}(1992)}]{vonDelft_Henley}%
  \BibitemOpen
  \bibfield  {author} {\bibinfo {author} {\bibfnamefont {J.}~\bibnamefont {von
  Delft}}\ and\ \bibinfo {author} {\bibfnamefont {C.~L.}\ \bibnamefont
  {Henley}},\ }\href {\doibase 10.1103/PhysRevLett.69.3236} {\bibfield
  {journal} {\bibinfo  {journal} {Phys. Rev. Lett.}\ }\textbf {\bibinfo
  {volume} {69}},\ \bibinfo {pages} {3236} (\bibinfo {year}
  {1992})}\BibitemShut {NoStop}%
\bibitem [{\citenamefont {C\'epas}\ and\ \citenamefont {Ralko}(2011)}]{Cepas}%
  \BibitemOpen
  \bibfield  {author} {\bibinfo {author} {\bibfnamefont {O.}~\bibnamefont
  {C\'epas}}\ and\ \bibinfo {author} {\bibfnamefont {A.}~\bibnamefont
  {Ralko}},\ }\href {\doibase 10.1103/PhysRevB.84.020413} {\bibfield  {journal}
  {\bibinfo  {journal} {Phys. Rev. B}\ }\textbf {\bibinfo {volume} {84}},\
  \bibinfo {pages} {020413} (\bibinfo {year} {2011})}\BibitemShut {NoStop}%
\bibitem [{\citenamefont {Castelnovo}\ \emph {et~al.}(2005)\citenamefont
  {Castelnovo}, \citenamefont {Chamon}, \citenamefont {Mudry},\ and\
  \citenamefont {Pujol}}]{Castelnovo_three_coloring}%
  \BibitemOpen
  \bibfield  {author} {\bibinfo {author} {\bibfnamefont {C.}~\bibnamefont
  {Castelnovo}}, \bibinfo {author} {\bibfnamefont {C.}~\bibnamefont {Chamon}},
  \bibinfo {author} {\bibfnamefont {C.}~\bibnamefont {Mudry}}, \ and\ \bibinfo
  {author} {\bibfnamefont {P.}~\bibnamefont {Pujol}},\ }\href {\doibase
  10.1103/PhysRevB.72.104405} {\bibfield  {journal} {\bibinfo  {journal} {Phys.
  Rev. B}\ }\textbf {\bibinfo {volume} {72}},\ \bibinfo {pages} {104405}
  (\bibinfo {year} {2005})}\BibitemShut {NoStop}%
\bibitem [{\citenamefont {Luttinger}\ and\ \citenamefont
  {Tisza}(1946)}]{Luttinger_Tisza}%
  \BibitemOpen
  \bibfield  {author} {\bibinfo {author} {\bibfnamefont {J.~M.}\ \bibnamefont
  {Luttinger}}\ and\ \bibinfo {author} {\bibfnamefont {L.}~\bibnamefont
  {Tisza}},\ }\href@noop {} {\bibfield  {journal} {\bibinfo  {journal} {Phys.
  Rev.}\ }\textbf {\bibinfo {volume} {70}},\ \bibinfo {pages} {954} (\bibinfo
  {year} {1946})}\BibitemShut {NoStop}%
\bibitem [{\citenamefont {Elser}(1989)}]{Elser_89}%
  \BibitemOpen
  \bibfield  {author} {\bibinfo {author} {\bibfnamefont {V.}~\bibnamefont
  {Elser}},\ }\href {\doibase 10.1103/PhysRevLett.62.2405} {\bibfield
  {journal} {\bibinfo  {journal} {Phys. Rev. Lett.}\ }\textbf {\bibinfo
  {volume} {62}},\ \bibinfo {pages} {2405} (\bibinfo {year}
  {1989})}\BibitemShut {NoStop}%
\bibitem [{\citenamefont {Domenge}\ \emph {et~al.}(2005)\citenamefont
  {Domenge}, \citenamefont {Sindzingre}, \citenamefont {Lhuillier},\ and\
  \citenamefont {Pierre}}]{Lhuillier_05}%
  \BibitemOpen
  \bibfield  {author} {\bibinfo {author} {\bibfnamefont {J.-C.}\ \bibnamefont
  {Domenge}}, \bibinfo {author} {\bibfnamefont {P.}~\bibnamefont {Sindzingre}},
  \bibinfo {author} {\bibfnamefont {C.}~\bibnamefont {Lhuillier}}, \ and\
  \bibinfo {author} {\bibfnamefont {L.}~\bibnamefont {Pierre}},\ }\href
  {\doibase 10.1103/PhysRevB.72.024433} {\bibfield  {journal} {\bibinfo
  {journal} {Phys. Rev. B}\ }\textbf {\bibinfo {volume} {72}},\ \bibinfo
  {pages} {024433} (\bibinfo {year} {2005})}\BibitemShut {NoStop}%
\bibitem [{\citenamefont {{Xiong}}\ and\ \citenamefont {{Wen}}(2012)}]{Wen_12}%
  \BibitemOpen
  \bibfield  {author} {\bibinfo {author} {\bibfnamefont {Z.}~\bibnamefont
  {{Xiong}}}\ and\ \bibinfo {author} {\bibfnamefont {X.-G.}\ \bibnamefont
  {{Wen}}},\ }\href@noop {} {\bibfield  {journal} {\bibinfo  {journal} {ArXiv
  e-prints}\ } (\bibinfo {year} {2012})},\ \Eprint
  {http://arxiv.org/abs/1208.1512} {arXiv:1208.1512 [cond-mat.stat-mech]}
  \BibitemShut {NoStop}%
\bibitem [{\citenamefont {Schulenburg}\ \emph {et~al.}(2002)\citenamefont
  {Schulenburg}, \citenamefont {Honecker}, \citenamefont {Schnack},
  \citenamefont {Richter},\ and\ \citenamefont {Schmidt}}]{Richter_magnons}%
  \BibitemOpen
  \bibfield  {author} {\bibinfo {author} {\bibfnamefont {J.}~\bibnamefont
  {Schulenburg}}, \bibinfo {author} {\bibfnamefont {A.}~\bibnamefont
  {Honecker}}, \bibinfo {author} {\bibfnamefont {J.}~\bibnamefont {Schnack}},
  \bibinfo {author} {\bibfnamefont {J.}~\bibnamefont {Richter}}, \ and\
  \bibinfo {author} {\bibfnamefont {H.-J.}\ \bibnamefont {Schmidt}},\ }\href
  {\doibase 10.1103/PhysRevLett.88.167207} {\bibfield  {journal} {\bibinfo
  {journal} {Phys. Rev. Lett.}\ }\textbf {\bibinfo {volume} {88}},\ \bibinfo
  {pages} {167207} (\bibinfo {year} {2002})}\BibitemShut {NoStop}%
\bibitem [{\citenamefont {Bergman}\ \emph {et~al.}(2008)\citenamefont
  {Bergman}, \citenamefont {Wu},\ and\ \citenamefont
  {Balents}}]{Bergman_Balents}%
  \BibitemOpen
  \bibfield  {author} {\bibinfo {author} {\bibfnamefont {D.~L.}\ \bibnamefont
  {Bergman}}, \bibinfo {author} {\bibfnamefont {C.}~\bibnamefont {Wu}}, \ and\
  \bibinfo {author} {\bibfnamefont {L.}~\bibnamefont {Balents}},\ }\href
  {\doibase 10.1103/PhysRevB.78.125104} {\bibfield  {journal} {\bibinfo
  {journal} {Phys. Rev. B}\ }\textbf {\bibinfo {volume} {78}},\ \bibinfo
  {pages} {125104} (\bibinfo {year} {2008})}\BibitemShut {NoStop}%
\bibitem [{\citenamefont {Zhitomirsky}\ and\ \citenamefont
  {Tsunetsugu}(2004)}]{Zhitomirsky_Tsunetsugu}%
  \BibitemOpen
  \bibfield  {author} {\bibinfo {author} {\bibfnamefont {M.~E.}\ \bibnamefont
  {Zhitomirsky}}\ and\ \bibinfo {author} {\bibfnamefont {H.}~\bibnamefont
  {Tsunetsugu}},\ }\href {\doibase 10.1103/PhysRevB.70.100403} {\bibfield
  {journal} {\bibinfo  {journal} {Phys. Rev. B}\ }\textbf {\bibinfo {volume}
  {70}},\ \bibinfo {pages} {100403} (\bibinfo {year} {2004})}\BibitemShut
  {NoStop}%
\bibitem [{\citenamefont {Richter}\ \emph {et~al.}(2004)\citenamefont
  {Richter}, \citenamefont {Schulenburg},\ and\ \citenamefont
  {Honecker}}]{Richter_lectures}%
  \BibitemOpen
  \bibfield  {author} {\bibinfo {author} {\bibfnamefont {J.}~\bibnamefont
  {Richter}}, \bibinfo {author} {\bibfnamefont {J.}~\bibnamefont
  {Schulenburg}}, \ and\ \bibinfo {author} {\bibfnamefont {A.}~\bibnamefont
  {Honecker}},\ }\enquote {\bibinfo {title} {Quantum magnetism in two
  dimensions: From semi-classical n{\'e}el order to magnetic disorder},}\ in\
  \href {\doibase 10.1007/BFb0119592} {\emph {\bibinfo {booktitle} {Quantum
  Magnetism}}},\ \bibinfo {editor} {edited by\ \bibinfo {editor} {\bibfnamefont
  {U.}~\bibnamefont {Schollw{\"o}ck}}, \bibinfo {editor} {\bibfnamefont
  {J.}~\bibnamefont {Richter}}, \bibinfo {editor} {\bibfnamefont {D.~J.~J.}\
  \bibnamefont {Farnell}}, \ and\ \bibinfo {editor} {\bibfnamefont {R.~F.}\
  \bibnamefont {Bishop}}}\ (\bibinfo  {publisher} {Springer Berlin
  Heidelberg},\ \bibinfo {address} {Berlin, Heidelberg},\ \bibinfo {year}
  {2004})\ pp.\ \bibinfo {pages} {85--153}\BibitemShut {NoStop}%
\bibitem [{\citenamefont {Schmidt}(2002)}]{Schmidt_2002}%
  \BibitemOpen
  \bibfield  {author} {\bibinfo {author} {\bibfnamefont {H.-J.}\ \bibnamefont
  {Schmidt}},\ }\href {http://stacks.iop.org/0305-4470/35/i=31/a=302}
  {\bibfield  {journal} {\bibinfo  {journal} {Journal of Physics A:
  Mathematical and General}\ }\textbf {\bibinfo {volume} {35}},\ \bibinfo
  {pages} {6545} (\bibinfo {year} {2002})}\BibitemShut {NoStop}%
\bibitem [{Note1()}]{Note1}%
  \BibitemOpen
  \bibinfo {note} {The $\protect \sqrt {3}\times \protect \sqrt {3}$ pattern
  can fit on cylinders of width that are multiples of 4 but not multiples of 6
  by shifting $\pm 2$ lattice constants in the periodic boundary condition
  along the cylinder circumference. These are $YC4+2$, $YC8-2$, $YC16-2$,
  $YC20+2$, etc. in the notation used in the DMRG literature. For cylinders of
  width that are multiples of 6, no such shifts are required}\BibitemShut
  {NoStop}%
\bibitem [{\citenamefont {Kumar}\ \emph {et~al.}(2014)\citenamefont {Kumar},
  \citenamefont {Sun},\ and\ \citenamefont {Fradkin}}]{Kumar2014}%
  \BibitemOpen
  \bibfield  {author} {\bibinfo {author} {\bibfnamefont {K.}~\bibnamefont
  {Kumar}}, \bibinfo {author} {\bibfnamefont {K.}~\bibnamefont {Sun}}, \ and\
  \bibinfo {author} {\bibfnamefont {E.}~\bibnamefont {Fradkin}},\ }\href
  {\doibase 10.1103/PhysRevB.90.174409} {\bibfield  {journal} {\bibinfo
  {journal} {Phys. Rev. B}\ }\textbf {\bibinfo {volume} {90}},\ \bibinfo
  {pages} {174409} (\bibinfo {year} {2014})}\BibitemShut {NoStop}%
\bibitem [{\citenamefont {Kumar}\ \emph {et~al.}(2016)\citenamefont {Kumar},
  \citenamefont {Changlani}, \citenamefont {Clark},\ and\ \citenamefont
  {Fradkin}}]{KumarChanglani2016}%
  \BibitemOpen
  \bibfield  {author} {\bibinfo {author} {\bibfnamefont {K.}~\bibnamefont
  {Kumar}}, \bibinfo {author} {\bibfnamefont {H.~J.}\ \bibnamefont
  {Changlani}}, \bibinfo {author} {\bibfnamefont {B.~K.}\ \bibnamefont
  {Clark}}, \ and\ \bibinfo {author} {\bibfnamefont {E.}~\bibnamefont
  {Fradkin}},\ }\href {\doibase 10.1103/PhysRevB.94.134410} {\bibfield
  {journal} {\bibinfo  {journal} {Phys. Rev. B}\ }\textbf {\bibinfo {volume}
  {94}},\ \bibinfo {pages} {134410} (\bibinfo {year} {2016})}\BibitemShut
  {NoStop}%
\bibitem [{\citenamefont {He}\ and\ \citenamefont {Chen}(2015)}]{He_PRL}%
  \BibitemOpen
  \bibfield  {author} {\bibinfo {author} {\bibfnamefont {Y.-C.}\ \bibnamefont
  {He}}\ and\ \bibinfo {author} {\bibfnamefont {Y.}~\bibnamefont {Chen}},\
  }\href {\doibase 10.1103/PhysRevLett.114.037201} {\bibfield  {journal}
  {\bibinfo  {journal} {Phys. Rev. Lett.}\ }\textbf {\bibinfo {volume} {114}},\
  \bibinfo {pages} {037201} (\bibinfo {year} {2015})}\BibitemShut {NoStop}%
\bibitem [{\citenamefont {{L{\"a}uchli}}\ \emph {et~al.}(2016)\citenamefont
  {{L{\"a}uchli}}, \citenamefont {{Sudan}},\ and\ \citenamefont
  {{Moessner}}}]{Lauchli_48}%
  \BibitemOpen
  \bibfield  {author} {\bibinfo {author} {\bibfnamefont {A.~M.}\ \bibnamefont
  {{L{\"a}uchli}}}, \bibinfo {author} {\bibfnamefont {J.}~\bibnamefont
  {{Sudan}}}, \ and\ \bibinfo {author} {\bibfnamefont {R.}~\bibnamefont
  {{Moessner}}},\ }\href@noop {} {\bibfield  {journal} {\bibinfo  {journal}
  {ArXiv e-prints}\ } (\bibinfo {year} {2016})},\ \Eprint
  {http://arxiv.org/abs/1611.06990} {arXiv:1611.06990 [cond-mat.str-el]}
  \BibitemShut {NoStop}%
\bibitem [{Ite()}]{Itensor}%
  \BibitemOpen
  \href@noop {} {}\bibinfo {note} {M. Stoudenmire and S.R. White,
  www.itensor.org}\BibitemShut {NoStop}%
\bibitem [{Note2()}]{Note2}%
  \BibitemOpen
  \bibinfo {note} {This argument does not account for the possibility that the
  $XXZ$ model for any $J_z\geq -1/2$ is potentially part of a critical
  line~\cite {Changlani_PRL}}\BibitemShut {NoStop}%
\bibitem [{\citenamefont {Changlani}\ \emph {et~al.}(2009)\citenamefont
  {Changlani}, \citenamefont {Kinder}, \citenamefont {Umrigar},\ and\
  \citenamefont {Chan}}]{Changlani_CPS}%
  \BibitemOpen
  \bibfield  {author} {\bibinfo {author} {\bibfnamefont {H.~J.}\ \bibnamefont
  {Changlani}}, \bibinfo {author} {\bibfnamefont {J.~M.}\ \bibnamefont
  {Kinder}}, \bibinfo {author} {\bibfnamefont {C.~J.}\ \bibnamefont {Umrigar}},
  \ and\ \bibinfo {author} {\bibfnamefont {G.~K.-L.}\ \bibnamefont {Chan}},\
  }\href {\doibase 10.1103/PhysRevB.80.245116} {\bibfield  {journal} {\bibinfo
  {journal} {Phys. Rev. B}\ }\textbf {\bibinfo {volume} {80}},\ \bibinfo
  {pages} {245116} (\bibinfo {year} {2009})}\BibitemShut {NoStop}%
\bibitem [{\citenamefont {Waldtmann}\ \emph {et~al.}(1998)\citenamefont
  {Waldtmann}, \citenamefont {Everts}, \citenamefont {Bernu}, \citenamefont
  {Lhuillier}, \citenamefont {Sindzingre}, \citenamefont {Lecheminant},\ and\
  \citenamefont {Pierre}}]{Waldtmann_Lhuillier}%
  \BibitemOpen
  \bibfield  {author} {\bibinfo {author} {\bibfnamefont {C.}~\bibnamefont
  {Waldtmann}}, \bibinfo {author} {\bibfnamefont {H.-U.}\ \bibnamefont
  {Everts}}, \bibinfo {author} {\bibfnamefont {B.}~\bibnamefont {Bernu}},
  \bibinfo {author} {\bibfnamefont {C.}~\bibnamefont {Lhuillier}}, \bibinfo
  {author} {\bibfnamefont {P.}~\bibnamefont {Sindzingre}}, \bibinfo {author}
  {\bibfnamefont {P.}~\bibnamefont {Lecheminant}}, \ and\ \bibinfo {author}
  {\bibfnamefont {L.}~\bibnamefont {Pierre}},\ }\href {\doibase
  10.1007/s100510050274} {\bibfield  {journal} {\bibinfo  {journal} {The
  European Physical Journal B - Condensed Matter and Complex Systems}\ }\textbf
  {\bibinfo {volume} {2}},\ \bibinfo {pages} {501} (\bibinfo {year}
  {1998})}\BibitemShut {NoStop}%
\bibitem [{\citenamefont {Mila}(1998)}]{Mila_kagome}%
  \BibitemOpen
  \bibfield  {author} {\bibinfo {author} {\bibfnamefont {F.}~\bibnamefont
  {Mila}},\ }\href {\doibase 10.1103/PhysRevLett.81.2356} {\bibfield  {journal}
  {\bibinfo  {journal} {Phys. Rev. Lett.}\ }\textbf {\bibinfo {volume} {81}},\
  \bibinfo {pages} {2356} (\bibinfo {year} {1998})}\BibitemShut {NoStop}%
\end{thebibliography}%

\end{document}